\documentclass[aps,prb,superscriptaddress,twocolumn]{revtex4}
\usepackage{bm}
\usepackage{graphicx}
\usepackage{amsfonts}
\usepackage{amsmath}

\begin{document}

\title{Quantum Heisenberg antiferromagnets:\\
 a survey of the activity in Firenze}

\author{Umberto Balucani}
\affiliation{Istituto dei Sistemi Complessi,
             Consiglio Nazionale delle Ricerche,
             via Madonna del Piano, I-50019 Sesto Fiorentino (FI), Italy}
\author{Luca Capriotti}
\affiliation{Valuation Risk Group, Credit Suisse First Boston (Europe) Ltd.,
             One Cabot Square, London E14 4QJ, United Kingdom}
\affiliation{Kavli Institute for Theoretical Physics, University
    of California, Santa Barbara, CA 93106, USA}
\author{Alessandro Cuccoli}
\affiliation{Dipartimento di Fisica dell'Universit\`a di Firenze,
             via G. Sansone 1, I-50019 Sesto Fiorentino (FI), Italy}
\affiliation{Istituto Nazionale per la Fisica della Materia,
             Unit\`a di Ricerca di Firenze,
             via G. Sansone 1, I-50019 Sesto Fiorentino (FI), Italy}
\author{Andrea Fubini}
\affiliation{Dipartimento di Fisica dell'Universit\`a di Firenze,
             via G. Sansone 1, I-50019 Sesto Fiorentino (FI), Italy}
\affiliation{Istituto Nazionale per la Fisica della Materia,
             Unit\`a di Ricerca di Firenze,
             via G. Sansone 1, I-50019 Sesto Fiorentino (FI), Italy}
\author{Tommaso Roscilde}
\affiliation{Department of Physics and Astronomy,
             University of Southern California,
             Los Angeles, CA 90089-0484, USA}
\author{Valerio Tognetti}
\affiliation{Dipartimento di Fisica dell'Universit\`a di Firenze,
             via G. Sansone 1, I-50019 Sesto Fiorentino (FI), Italy}
\affiliation{Istituto Nazionale per la Fisica della Materia,
             Unit\`a di Ricerca di Firenze,
             via G. Sansone 1, I-50019 Sesto Fiorentino (FI), Italy}
\author{Ruggero Vaia}
\affiliation{Istituto dei Sistemi Complessi,
             Consiglio Nazionale delle Ricerche,
             via Madonna del Piano, I-50019 Sesto Fiorentino (FI), Italy}
\affiliation{Istituto Nazionale per la Fisica della Materia,
             Unit\`a di Ricerca di Firenze,
             via G. Sansone 1, I-50019 Sesto Fiorentino (FI), Italy}
\author{Paola Verrucchi}
\affiliation{Istituto Nazionale per la Fisica della Materia,
             Unit\`a di Ricerca di Firenze,
             via G. Sansone 1, I-50019 Sesto Fiorentino (FI), Italy}
\affiliation{Istituto dei Sistemi Complessi,
             Consiglio Nazionale delle Ricerche,
             via Madonna del Piano, I-50019 Sesto Fiorentino (FI), Italy}

\date{\today}

\begin{abstract}
Over the years the research group in Firenze has produced a number of
theoretical results concerning the statistical mechanics of quantum
antiferromagnetic models, which range from the theory of two-magnon
Raman scattering to the characterization of the phase transitions in
quantum low-dimensional antiferromagnetic models. Our research
activity was steadily aimed to the understanding of experimental
observations.
\end{abstract}


\maketitle

\section{Introduction}

The Heisenberg model may well be considered the cornerstone of the
modern theory of magnetic systems; the reason for such an important
role is the simple structure of the Hamiltonian, whose symmetries
underlie its peculiar features. The basic forces which determine the
alignment of the spins, are represented by the exchange integrals
$J$'s. At variance with the ferromagnet where the parallel alignment
is promoted, in the antiferromagnet a lot of peculiar arrangements of
the spins can occur, with strong differences between classical and
quantum systems. As matter of fact, also for nearest-neighbor
antiferromagnetic interactions the ground state of the Hamiltonian is
different from the N\'eel state with antialigned spins, and the
(staggered) magnetization shows the so called {\em spin reduction}
with respect to the saturation value also at $T\,{=}\,0$. The linear
excitations of an antiferromagnet can be roughly associated in two
families and pair excitations with vanishing total magnetization are
possible: the fact that the total momentum of these can be close to
zero allows for their investigation by light scattering.

While these peculiar features of antiferromagnetism already occur in
three-dimensional (3D) compounds, they are more pronounced in the
low-dimensional ones, where other effects caused by the enhanced role
of classical and quantum fluctuations are present and exotic spin
configurations associated with field theory models can appear.
Indeed, the last two decades have seen a renewed interest both in the
case of the one-dimensional (1D) quantum Heisenberg antiferromagnet
(QHAF), for which a peculiar behavior of the ground state vs spin
value was predicted~\cite{Haldane1983}, and of the two-dimensional
(2D) QHAF, because of its theoretically challenging properties and of
the fact that it models the magnetic behavior of the parent compounds
of some high-$T_{\rm{c}}$
superconductors~\cite{SokolP1993,Chakravarty1990}. The experimental
activity on 2D antiferromagnets stems from the existence of several
real compounds whose crystal structure is such that the magnetic ions
form parallel planes and interact strongly only if belonging to the
same plane. As a consequence of such structure, their magnetic
behavior is indeed 2D down to those low temperatures where the weak
interplane interaction becomes relevant, driving the system towards a
3D ordered phase.

In addition, the 2D Heisenberg model can be enriched trough
symmetry-breaking terms -- we considered easy-axis (EA) and
easy-plane (EP) anisotropy, as well as an external uniform magnetic
field -- which are useful to reproduce the experimental behavior of
many layered compounds. In the EA case one is left with a discrete
reflection symmetry and the system undergoes an Ising-like phase
transition. In the EP case or when a magnetic field is applied the
residual $O(2)$ symmetry prevents finite-temperature
ordering~\cite{MerminW1966}, but vortex excitations are possible and
determine a Berezinskii-Kosterlitz-Thouless (BKT) transition between
a paramagnetic and a quasi-ordered phase. In spite of the tiny
anisotropies of real systems (usually $\lesssim{0.01}\,J$), it can be
shown that they dramatically change the behavior of the spin array
already at temperatures of the order of $J$.

In this paper we report about the progresses in the theory of
Heisenberg antiferromagnets that have been obtained by our group in
Firenze. The early work on the theory of two-magnon Raman scattering
is summarized in Section~\ref{s.2magnon}, while the following
Sections report about the recent activity on low-dimensional
antiferromagnetism. Section~\ref{s.1D} is devoted to 1D models, and
concerns the study of the effect of soliton-like excitations in the
compound TMMC, as well as the anisotropic spin-$1$ model, for which a
reduced description of the ground state allows one to investigate the
quantum phase transition in a unitarily transformed representation
and to obtain quantitative results for the phase diagram.
Section~\ref{s.isotropic} concerns the theory of the isotropic 2D
QHAF, for which we reproduced the experimental correlation length by
means of a semiclassical approach, also deriving the connection with
(and the limitations of) famous quantum field theory results. In
Section~\ref{s.anisotropic} we summarize several recent results
concerning the anisotropic 2D QHAF, with emphasis onto the different
phase diagrams and the experimentally measurable signatures of XY or
Ising behavior. Eventually, in Section~\ref{s.J1J2} results on the 2D
frustrated $J_1$-$J_2$ isotropic model are described.

\section{Two-magnon Raman scattering in Heisenberg antiferromagnets}
\label{s.2magnon}

The scattering of radiation is a very powerful tool to study
elementary excitations in Condensed Matter Physics. Any complete
experiment gives rise to a quasi-elastic component due to
non-propagating or diffusive modes and to symmetrically shifted
spectra corresponding to the states of the system under investigation
with an amplitude ratio governed by the detailed balance principle.
The most sensitive probes for this investigation are undoubtedly
thermal neutrons, because the characteristic energies and wavevectors
fit very well with those of the magnetic elementary excitations.
However, light-scattering experiments can require a simpler apparatus
and offer a better accuracy, although the transfer wavevector
${\bm{k}}$ is much smaller than the size of the Brillouin zone so
that usually only the center of this zone can be directly probed. In
spite of this, two-spin Raman scattering involving the creation and
destruction of a pair of elementary excitations can be performed,
with the contribution of two magnons having equal frequencies and
opposite wavevectors. This two-magnon scattering is expected to be
spread over a band of frequencies in antiferromagnets. However, the
density of states strongly enhances the contribution of zone-boundary
(ZB) excitations~\cite{Fleury1970}, i.e., at
${\bm{k}}\,{\sim}\,{\bm{k}}_{\rm{ZB}}$.

While in ferromagnets the two-spin process is only due to a second
order mechanism, orders of magnitude smaller than the first order
one, in antiferromagnets a different independent process is
permitted, stronger than the corresponding one for single-spin
spectra~\cite{Loudon1968}. Specifically, an exchange mechanism does
not change the total $z$-component of the spins: exciting two magnons
in the two different sublattices
$(\Delta{M}\,{=}\,0)$~\cite{Fleury1969} is the dominant scattering
process.

The one-spin Raman scattering peak disappears at the N\'eel
temperature because it probes the smallest wavevectors, related with
the long-range correlations. In contrast, two-magnon Raman scattering
essentially probes the highest wavevectors, related to short-range
correlations. Therefore two-spin Raman scattering features persist
also in the paramagnetic phase~\cite{Fleury1969} where short-range
order is still present.

Let us consider the following antiferromagnetic Hamiltonian with
exchange integral $J\,{>}\,0$, $z$ nearest neighbors with
displacements labeled by $\bm{d}$, and two $(a,b)$
sublattices~\cite{BT1976nc}:
\begin{equation}\label{AFM}
 {\cal H}=\frac J2\sum_{\bm{id}}\,
 {\bm{S}}_{{\bm{i}},a}{\cdot}{\bm{S}}_{\bm{i+d},b}\,.
\end{equation}
The scattering cross section ${\cal S}(\omega)$ turns
out~\cite{ElliottT1969} to be proportional to the Fourier transform
of $\langle{M(0)M(t)}\rangle$, where
\begin{equation}
 \label{Mj}
 M=\sum_{\bm{k}} {\cal M}_{\bm{k}}~
 {\bm{S}}_{\bm{k}}{\cdot} {\bm{S}}_{-\bm{k}}\,,
\end{equation}
is the effective Raman scattering operator.

Many antiferromagnetic compounds can be mapped onto this model, even
though a small next-nearest-neighbor exchange interaction without
competitive effects, as well as anisotropy terms could be present.
For instance, there are 3D perovskite and rutile structures (e.g.,
KNiF$_4$, NiF$_2$) and 2D layered structures (e.g., K$_2$NiF$_4$,
LaCuO$_2$).

Let us remember that the exact ground state is not exactly known,
except in 1D models with $S\,{=}\,1/2$ or $S\,{=}\,\infty$ (i.e., the
classical case): in the latter case it coincides with the `N\'eel
state' with antialigned sublattices.

In the ordered phase the theory can be developed in terms of two
families of magnon operators
$(\alpha_{\bm{k}}\,,\,\beta_{\bm{k}}\,)$, through the Dyson-Maleev
spin-boson transformation and a Bogoliubov transformation:
\begin{equation}\label{BAFM}
{\cal H}=E_0+{\cal H}_0+V\,,
\end{equation}
where $E_0$ is the ground state energy in interacting spin-wave
approximation and
\begin{equation}\label{SW}
 {\cal H}_0=\sum_{\bm{k}} \omega_{\bm{k}}
 (\alpha_{\bm{k}}^\dag\alpha_{\bm{k}}+\beta_{\bm{k}}^\dag\beta_{\bm{k}})
\end{equation}
is the quadratic part of the Hamiltonian of a magnon gas whose
frequencies, renormalized by zero-$T$ quantum fluctuations, are
\begin{equation}\label{omega}
 \omega_{\bm{k}}=JSz\Big(1+\frac C{2S}\Big)
 \sqrt {1-\gamma^2_{\bm{k}}}\,;
 ~~~~~\gamma_{\bm{k}}=
 \frac 1z\sum_{\bm d} e^{-i{\bm{k}}{\cdot}\bm d}~.
\end{equation}

The last term in the Hamiltonian, $V$, represents the four-magnon
interaction, whose most significant term refers to two magnons of
each family and turns out to be:
\begin{equation}\label{Vint}
 V = 2\frac{Jz}N\sum_{\bm q\bm q'\bm p\bm p'}
 \delta_{\bm q+\bm p,\bm q'+\bm p'}\,
 I^{\alpha\beta}_{\bm q\bm q',\bm p\bm p'}
 \alpha_{\bm q}^\dag\alpha_{\bm q'}\beta_{\bm p}^\dag\beta_{\bm p'}\,,
\end{equation}
where the coefficients $I^{\alpha\beta}_{\bm{qq',pp'}}$ are known
functions of $\gamma_{\bm{k}}$.

In the Hartree-Fock approximation~\cite{BT1973prb} the temperature
dependent Raman scattering operator~(\ref{Mj}) can be written
\begin{equation}\label{HFMj}
 M=\alpha(T)~S\sum_{\bm{k}}\,\Phi_{\bm{k}}(\alpha_{\bm{k}}
 \beta_{\bm{k}}+\alpha^\dag_{\bm{k}}\beta^\dag_{\bm{k}})\,,
\end{equation}
with $\omega_{\bm{k}}(T)\,{=}\,\alpha(T)\,\omega_{\bm{k}}$. The two
magnons created or destroyed by the operator~(\ref{HFMj}) interact
through $V$ as given by~(\ref{Vint}), so that the peak of the cross
section ${\cal S}(\omega)$ appears at values smaller that
$2\omega_{_{\rm{ZB}}}$ for an amount of the order of $J$. The
explicit ${\cal S}(\omega)$ at $T\,{=}\,0$ was calculated in the
`ladder approximation' by Elliott and Thorpe and found in very good
agreement with experiments~\cite{ElliottT1969}.

\begin{figure}
\includegraphics[width=85mm,angle=0]{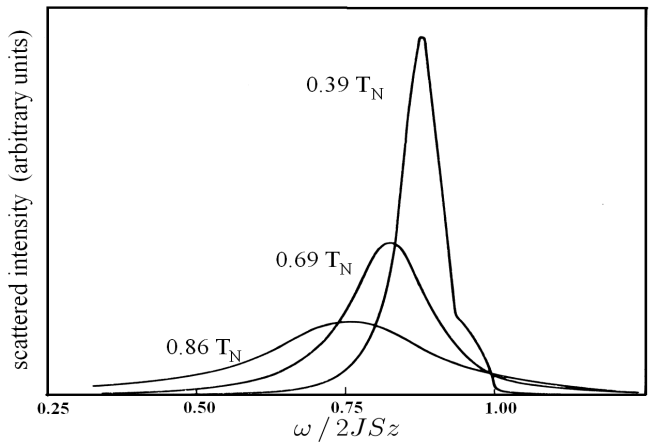}
\caption{\label{f.KNiFT}
Theoretical two-magnon spectra in KNiF$_3$ at different
temperatures~\cite{BT1973prb}. }
\end{figure}

\begin{figure}
\includegraphics[width=85mm,angle=0]{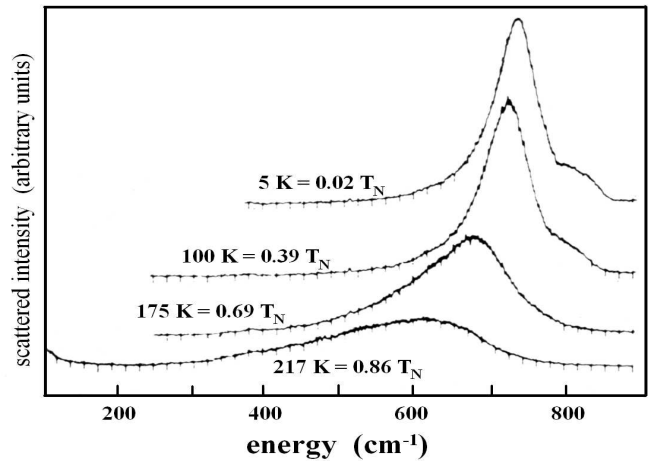}
\caption{\label{f.KNiFS}
Experimental two-magnon spectra in KNiF$_3$ at different
temperatures~\cite{Fleury1970}. }
\end{figure}

\begin{figure}
\includegraphics[width=85mm,angle=0]{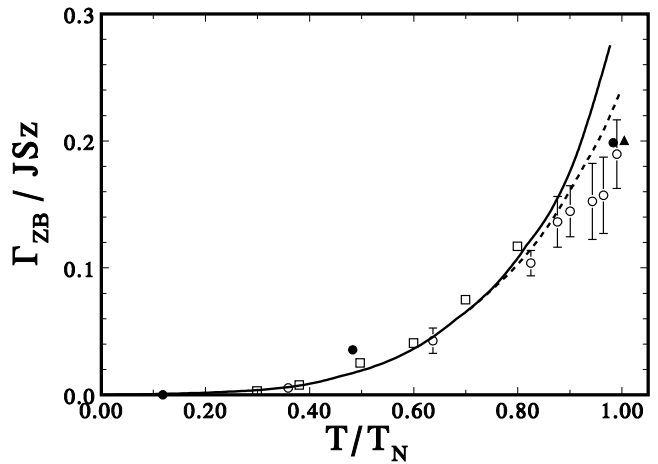}
\caption{\label{f.Rbdamp}
Zone-boundary damping $\Gamma_{\rm{ZB}}$ vs temperature. The symbols
refer to different experimental approaches: in particular the open
circles are our light scattering data~\cite{BMTZ1978ssc}. The dashed
line is an improvement~\cite{BMTZ1978ssc} to a previous (solid)
theoretical curve~\cite{BT1973prb}.}
\end{figure}

The finite temperature calculation of the two-magnon Raman scattering
cross section in the ordered region, up to
$T\,{\sim}\,0.95\,T_{\rm{N}}$ was performed by Balucani and
Tognetti~\cite{BT1973prb}, calculating the two-magnon propagator in
the `ladder approximation', taking also into account the damping and
the temperature renormalization of the magnons at the boundary of the
Brillouin zone~\cite{BLT2003phrep}. The calculated spectra ${\cal
S}(\omega)$, at increasing temperatures, were found in very good
agreement with the experimental ones~\cite{BT1976nc} and their
characteristic parameters (peak and line-width) permitted to
determine the temperature behavior of the frequency and damping of
the ZB magnons~\cite{BMTZ1978ssc}. In Figs.~\ref{f.KNiFT}
and~\ref{f.KNiFS} we show the excellent agreement of our theoretical
approach with the experiments in the ordered phase~\cite{CDZ1972}.
The validity of light scattering in probing the characteristic of ZB
magnons has been confirmed both from the theoretical and the
experimental point of view~\cite{BMTZ1978ssc}. In Fig.~\ref{f.Rbdamp}
our theoretical ZB magnon damping calculations are compared with
experimental data from different techniques.

In the paramagnetic phase all experimental spectra show the
persistence of a broad inelastic peak up to $T\,{\sim}\,1.4\,T_N$.
Only at $T\,{\gg}\,T_N$ the spectra have a structureless shape
centered around $\omega\,{=}\,0$. As matter of fact, the highest
wavevectors sample only the behavior of clusters of neighboring
spins, thus giving a measure of the short-range antiferromagnetic
order that is present at all finite temperatures.

In the disordered phase conventional many-body methods are of little
use for a quantitative interpretation of the observed largely spread
spectra. The concept of quasi-particle loses its meaning because of
the overdamped character of the `excitations'. The calculation of
${\cal S}(\omega)$ can be instead approached by other more general
theoretical methods devoted to the representation of the dynamical
correlation functions based on the linear response
theory~\cite{BLT2003phrep}. Let us consider the `Kubo relaxation
function' associated with our scattering process:
\begin{equation}\label{Kubo1}
 f_0(t)\equiv \frac 1{\langle M(0)M(0)\rangle}
 \int^\beta_0\!\!d\lambda\, \big\langle \,
 e^{\lambda{\cal H}}M(0)e^{-\lambda{\cal H}}M(t)\big\rangle \,.
\end{equation}
Its Laplace transform $f_0(z)$ is related to the scattering cross
section:
\begin{equation}\label{Kubo2}
 {\cal S}(\omega)\propto\frac\omega{1-e^{-\beta\omega}}
 ~\Re\,f_0(z\,{=}\,i\omega)\,.
\end{equation}
Mori~\cite{Mori1965} has given the following continued fraction
representation of the relaxation function~\cite{BLT2003phrep}:
\begin{equation}\label{Mori1}
 f_0(z)=\frac 1{z+\Delta_1f_1(z)}\,;
 ~~f_n(z)=\frac1{z+\Delta_{n+1}f_{n+1}(z)}\,,
\end{equation}
which is formally exact, but allow us to do some approximations about
the level of the termination $f_{n+1}(z)$. The quantities $\Delta_n$
can be expressed in terms of frequency moments:
\begin{equation}\label{Mori2}
 \langle\omega^{2n}\rangle
 = \int_{-\infty}^{\infty}d\omega\,\omega^{2n}f_0(\omega)\,.
\end{equation}
In our calculations of $f_0(z)$ in the entire paramagnetic
phase~\cite{BT1977prb,BPT1978nc}, the coefficients $\Delta_1$ e
$\Delta_2$ have been approximately evaluated by means of a decoupling
procedure. Moreover, the third stage of the continued
fraction~(\ref{Mori1}) is evaluated assuming that
\begin{equation}\label{f3}
 \Delta_3f_3(z)\sim\Delta_3[f_3(0)+zf_3'(0)]\,.
\end{equation}
The parameters involved in~(\ref{f3}) can be estimated by the
knowledge of the short time behavior of $f_0(t)$ determined by the
first moments, $\langle\omega^{2}\rangle$ and
$\langle\omega^{4}\rangle$.

\begin{figure}
\includegraphics[width=85mm,angle=0]{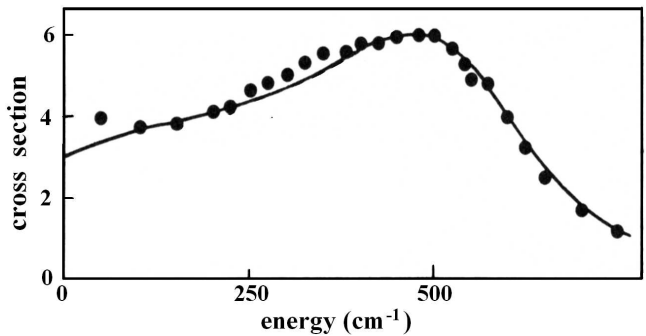}
\caption{\label{f.KNiFpara}
Two-spin Stokes spectrum in KNiF$_3$ at
$T\,{\simeq}\,{1.02}\,T_{\rm{N}}$. The line reports the theoretical
shape~\cite{BT1977prb}, compared with experimental data. }
\end{figure}

The results of our approach in the paramagnetic region are compared
with the experiment in Fig.~\ref{f.KNiFpara}, showing the persistence
of the peak of the ZB magnetic excitations above the critical
temperature.

\section{The one-dimensional antiferromagnet}
\label{s.1D}

\subsection{Solitons in the antiferromagnet TMMC}

Interest in low-dimensional systems is motivated by the much greater
simplicity of calculation as compared with the 3D ones. The powerful
mathematical approach based on the inverse-scattering and Bethe
Ansatz techniques permits to exactly solve some 1D models,
calculating thermodynamic and sometimes transport quantities both in
classical and quantum cases~\cite{KorepinBI1993}. The most celebrated
realizations of these models occur in 1D magnets. An original
suggestion by Mikeska~\cite{Mikeska1980} was that the
antiferromagnetic chain TMMC [(CH$_3$)$_4$NMnCl$_3$] can be mapped
onto a sine-Gordon classical 1D model. The elementary excitations of
the sine-Gordon field are given in terms of linear small-amplitude
spin-waves and non-linear breathers and kink-solitons. The non-linear
elementary excitations give a detectable contribution to the magnetic
specific heat.

TMMC is composed of Heisenberg ($S\,{=}\,5/2$) antiferromagnetic
chains along the $z$-axis:
\begin{equation}\label{TMMC1}
 {\cal H} = J\sum_i\big({\bm{S}}_i{\cdot}{\bm{S}}_{i+1}
 -\delta S^z_iS^z_{i+1}\big)\,,
\end{equation}
with a very small easy-plane anisotropy ($\delta\,{=}\,0.0086$).

A magnetic field of the order of $1\div10$\,T can be applied
perpendicularly ($y$-axis) or along the chain . In the first case,
with approximations the more valid the lower the magnetic field
($H\,{<}\,5$\,T), in the continuum limit TMMC can be represented by
the classical sine-Gordon Hamiltonian:
\begin{equation}\label{TMMC2}
 {\cal H}=\frac A2 \int dx\big[\dot{\Phi}^2+c_0^2\Phi_x^2
 +2 \omega_0^2(1-\cos\Phi)\big]\,,
\end{equation}
whose parameters are related with the magnetic
Hamiltonian~(\ref{TMMC1}), the reduced magnetic field
$h\,{=}\,g\mu_{\rm{B}}H$, and the lattice spacing $a$ as follows:
\begin{equation}\label{TMMC3}
 A=\frac 1{8Ja}\,,
 ~~c_0=aJS\sqrt{1{-}\frac\delta 2}\,,
 ~~\omega_0=h\sqrt{1{-}\frac\delta 2}~.
\end{equation}
The energy of kink-soliton turns out to be
\begin{equation}
E_s=8A\omega_0 c_0 \simeq h\,S~,
\end{equation}
and depends on the applied field. At difference with the
ferromagnetic solitons, these solitons can be easily excited at
lowest temperatures and can give a significant contribution to the
thermodynamics~\cite{CurrieKBT1980}. When the field is applied
longitudinally along the $z$-axis only spin-waves are present:
therefore, the specific-heat measurements were performed in the two
configurations. The contribution from the nonlinear excitations was
obtained as the difference $\Delta C$ between the two experiments.

\begin{figure}
\includegraphics[width=85mm,angle=0]{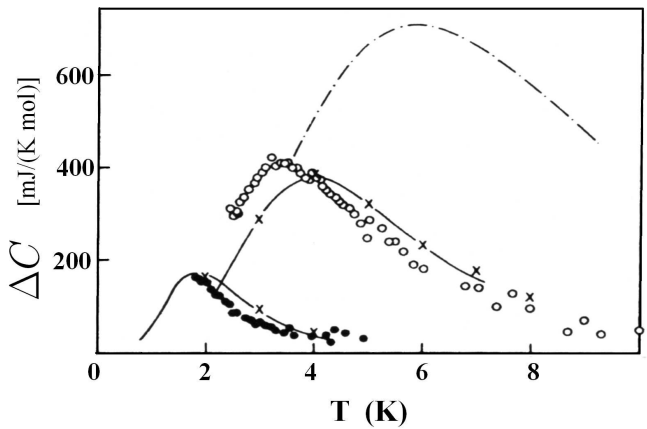}
\caption{\label{f.TMMC}
Experimental contribution of nonlinear excitations to the specific
heat of TMMC,
$\Delta{C}\,{=}\,C(H)\,{-}\,C(0)\,{-}\,\Delta{C}_{\rm{SW}}$. The
field values are $H\,{=}\,5.39$\,T ($\bullet$) and $H\,{=}\,2.5$\,T
($\circ$). The dash-dotted line reports the result of the free
soliton gas phenomenology. The planar model (interpolated crosses)
appears to quantitatively explain the behavior of TMMC. }
\end{figure}

The thermodynamic quantities were calculated by the classical
transfer-matrix method~\cite{SchneiderS1980} for the sine-Gordon
model~(\ref{TMMC2}). We then used a classical discrete planar
model~\cite{BPRT1983prb}:
\begin{equation}\label{TMMC4}
{\cal H}=
\sum_i\big[2JS^2\cos(\Phi_i-\Phi_{i+1})+hS(1-\cos\Phi_i)\big]\,,
\end{equation}
verifying that it is qualitatively similar to the sine-Gordon. The
comparison~\cite{BPRT1983prb} is shown in Fig.~\ref{f.TMMC}, where
the linear spin-wave specific heat was subtracted to emphasize the
nonlinear contribution, together with the prediction of the
`classical soliton gas phenomenology'~\cite{CurrieKBT1980}.

This proved the presence of nonlinear excitations similar to
sine-Gordon solitons, but the peak of the specific heat occurs at
temperatures where solitons cannot be considered to be non
interacting and the `classical soliton-gas phenomenology' breaks
down. When the magnetic field is increased up to $9.98\,T$ the
model~(\ref{TMMC4}) is no more able to describe the experiments. A
quasi-uniaxial model~\cite{BPRT1983prb} was proposed and found in
good agreement. For general reference on the subject
see~\cite{LBBT1984book}.

\subsection{The $S\,{=}\,1$ quantum antiferromagnet}
\label{s.Haldane}

We here deal with quantum antiferromagnetic spin chains, focusing our
attention on the class of models defined by the Hamiltonian
\begin{equation}
\frac{\cal{H}}{J}=\sum_i\Big[(S^x_i S^x_{i+1}{+}S^y_i S^y_{i+1}){+}
     \lambda S^z_i S^z_{i+1}{+}d(S^z_i)^2\Big]
\label{h.haldane}
\end{equation}
with exchange integral $J\,{>}\,0$ and single-ion anisotropy $d$.

One of the most surprising evidence of the difference between ferro-
and antiferromagnetic systems is related with the so-called Haldane
conjecture, i.e. with $T\,{=}\,0$ properties of integer-spin
antiferromagnetic chains. In general, we expect three possible
situations for the ground state of a magnetic system: either it is
ordered (with finitely constant correlation functions), or
quasi-ordered (with power-law decaying correlation functions), or
completely disordered (with exponentially decaying correlation
functions). One could intuitively expect the third option to be
possibly dismissed, based on the idea that, when thermal fluctuations
are completely suppressed, the system be in an ordered or at least
quasi-ordered ground state. This idea is in fact proved correct for
half-integer spin systems, thanks to the so called {\em
Lieb-Schultz-Mattis} theorem~\cite{LiebSM1961}. Despite the
generalization of such theorem to integer-spin systems being
impossible, its general validity has been taken for granted till
1983, when Haldane~\cite{Haldane1983} suggested, for the integer-spin
Heisenberg chain, an unexpected $T\,{=}\,0$ behavior: a unique and
genuinely disordered ground state, meaning exponentially decaying
correlation functions and a finite gap in the excitation spectrum.
After more than two decades Haldane's idea that integer-spin systems
can have a genuinely disordered ground state still stands as a
conjecture. However,
theoretical~\cite{AffleckKLT8788,Affleck1989,ArovasAH1988,Takahashi1989},
experimental~\cite{BuyersMAHGH1986,SteinerKKPP1987,DateK1990,BrunelBZBR1992,GaveauBRR1995}
and
numerical~\cite{BotetJK1983,ParkinsonB1985,NightingaleB1986,Kennedy1990,SakaiT1990,DelicaKLM1991,KoehlerS1992}
works have definitely confirmed its validity.

Let us consider Eq.~(\ref{h.haldane}) for integer spin: in the
$(d,\lambda)$ plane one may identify different quantum phases,
corresponding to models whose ground states share common features.
For $\lambda\,{>}\,0$ three phases are singled out: the N\'eel phase
($\lambda\,{\gg}\,d$), where the ground state has a N\'eel-like
structure, the so-called {\em large-}$d$ phase ($d\,{\gg}\,\lambda$),
where the ground state is characterized by a large majority of sites
where $S^z\,{=}\,0$, and the Haldane phase, which extends around the
isotropic point ($d\,{=}\,0,~\lambda\,{=}\,1$), and is characterized
by disordered ground states.

We first deal with the Ising limit,
${\cal{H}}/J\,{=}\,\lambda\sum_{i}S_{i}^{z}S_{i+1}^{z}~$: Upon its
ground state, the antiferromagnetically ordered N\'eel state, one may
construct three types of excitations: a single deviation, a direct
soliton, an indirect soliton, where {\em direct} ({\em indirect})
refers to the fact that the excitation be generated by flipping all
the spins on the right of a given site while keeping the $z$
component of the spin on such site unchanged (setting it to zero).
The above configurations have all energy ${+}2\lambda$ with respect
to that of the ground state, and do generate, when properly combined,
all the excited states; amongst them, we concentrate upon those
containing a couple of adjacent indirect solitons and notice that
their energy is ${+}3\lambda$, while excited states containing two
separate indirect solitons have energy ${+}4\lambda$. Therefore,
indirect solitons are characterized by a bounding energy $\lambda$;
moreover, one may easily see that isolated solitons may effectively
introduce disorder in the global configuration of the system, while
coupled solitons do only reduce the magnetization of each of the two
antiferromagnetic sublattices~\cite{Mikeska1995}. In fact, strings
containing any odd (even) number of adjacent solitons act on the
order of the global configuration as if they were isolated (coupled)
solitons.

As we move from the Ising limit, the transverse interaction
$\sum_{i}(S^x_iS^x_{i+1}+S_i^yS_{i+1}^y)$ comes into play, and is
seen~\cite{Mikeska1992} to more efficiently lower the energy of the
system by delocalizing indirect solitons rather than single
deviations or direct solitons, thus indicating configurations which
uniquely contain indirect solitons as crucial in understanding how
the system evolves from the Ising limit (N\'eel phase) to the
isotropic case(Haldane phase).

From the above ideas we may draw a simple but suggestive scheme for
such evolution:\par
\noindent - in the Ising limit ($\lambda\,{\rightarrow}\,\infty$)
the ground state is the antiferromagnetically ordered N\'eel state;
\par
\noindent - as $\lambda$ decreases, indirect solitons appear along the
chain in pairs, thus keeping the antiferromagnetic order;
\par\noindent - as $\lambda$ is further lowered, indirect
soliton pairs dissociate due to the transverse interaction which, by
spreading solitons along the chain, can cause the ground state to be
disordered.

Due to the privileged role of indirect solitons in the above scheme,
we concentrate on configurations which do only contain indirect
solitons. Such configurations generate a subspace for the Hilbert
space of the system, which is referred to, in the literature, as the
{\em reduced} Hilbert space~\cite{GomezSantos1989}. States belonging
to the {\em reduced} Hilbert space are strongly characterized by the
fact that if one eliminates all sites with $S^z\,{=}\,0$, a perfectly
antiferromagnetically ordered chain is left. Remarkably, this type of
order, which is called {\em hidden order} in the literature, is not
destroyed by soliton pairs dissociation, and it actually
characterizes the disordered ground state of a Haldane system, as
discussed below.

In 1992 Kennedy and Tasaki (KT) defined a non-local unitary
transformation~\cite{KennedyT1992} which makes the {\em hidden order}
visible, meanwhile clarifying its meaning. The transformation is
defined by $U\,{=}\,({-}1)^{N_{0}{+}[N/2]}\prod_k U_{k}$ with
\[
 U_k=
 \frac{1}{2}\Big(e^{i\pi\sum_{p=1}^{k-1}S_{p}^{z}}{-}1\Big)
 e^{i\pi{S_{k}^{x}}}+
 \frac{1}{2}\Big(e^{i\pi\sum_{p=1}^{k-1}S_{p}^{z}}{+}1\Big)~,
\]
where $N$ is the number of sites of the chain, $[N/2]$ is the integer
part of $N/2$, and $N_0$ is the number of odd sites where
$S^z\,{=}\,0$. If the pure state $|\Psi\rangle$ has {\em hidden}
order, meaning that it only contains indirect solitons, then
$U|\Psi\rangle$ has spins with $S^{z}\,{\neq}\,0$ all parallel to
each other. This point is made transparent by the introduction of the
{\em string} order parameter~\cite{denNijsR1989}
\begin{equation}
    O_{\rm{string}}^{\alpha}({\cal H})
    \equiv\lim_{|i-j|\rightarrow \infty}
    \bigg\langle
S^{\alpha}_i\,\exp\bigg[i\pi\sum_{l=i}^{j-1}S^{\alpha}_l\bigg]
    \, S^{\alpha}_j\bigg\rangle_{\cal H}~,
  \label{sop}
\end{equation}
where $\alpha\,{=}\,x,y,z$, and $\langle\cdots\rangle_{\cal{H}}$
indicates the expectation value over the ground state of the
Hamiltonian ${\cal{H}}$. It may be shown that
$O_{\rm{string}}^z({\cal H})\,{\neq}\,0$ if and only if the ground
state belongs to the {\em reduced} Hilbert space. In other terms,
while ferromagnetic order is revealed by the ferromagnetic order
parameter
$O^{\alpha}_{\rm{ferro}}\,{\equiv}\,\lim_{|i-j|\rightarrow\infty}
\langle S^{\alpha}_i S^{\alpha}_j\rangle_{\cal{H}}$\,,
the {\em hidden} order is revealed by the {\em string} order
parameter Eq.~(\ref{sop}). In fact, the non local transformation $U$
relates the above order parameters through the relation
\begin{equation}
{O^{\eta}_{\rm{string}}}=O^{\eta}_{\rm{ferro}}(U{\cal{H}} U^{-1})~,
\label{O.string-O.ferro}
\end{equation}
for $\eta\,{=}\,x,z$, meaning that the analysis of the {\em hidden}
order in a system described by ${\cal{H}}$ may be developed by
studying the ferromagnetic order in the system described by the
transformed Hamiltonian
${\widetilde{\cal{H}}}{\equiv}U{\cal{H}}U^{-1}$, which reads, for
${\cal{H}}$ defined by Eq.~(\ref{h.haldane}),
\begin{equation}
\frac{\widetilde{\cal{H}}}{J}\,{=}\,\sum_{i}\Big[\!-S_{i}^{x}S_{i+1}^{x}{+}
S_{i}^{y}e^{S_{i}^{z}+S_{i+1}^{x}}S_{i+1}^{y}{-}
\lambda S_{i}^{z}S_{i+1}^{z}{+}d (S_{i}^{z})^2\!\Big] .
\label{h.tilde}
\end{equation}

Our work developed as follows: one first assumes that the relevant
configurations, as far as the N\'eel-Haldane transition is concerned,
belong to the {\em reduced} Hilbert space; this permits, by the KT
transformation, to restrict the analysis to the subspace of states
with either $S^z_i\,{=}\,1$ or $S^z_i\,{=}\,0$, $\forall i$. Then the
expectation value of the transformed Hamiltonian
$\widetilde{\cal{H}}$ Eq.(\ref{h.tilde}) is minimized on a trial
ground state whose structure takes into account at least short-range
correlations between spins. By this procedure, we aim at following
the effective dissociation of soliton pairs, in order to clarify the
connection between the occurrence of isolated solitons in the ground
state, and the transition towards the completely disordered
Haldane-phase~\cite{KoehlerS1992,Mikeska1992,GomezSantos1989,Tasaki1991}.

In the framework of a standard variational approach, we should
minimize $\langle\Phi_0|\,{\cal H}\,|\Phi_0\rangle$ with respect to a
certain number of variational parameters entering the expression of
the normalized trial ground-state $|\Phi_0\rangle$. By applying the
non-local unitary transformation ${\cal U}$ we instead minimize
$\langle\Psi_0|{\cal{U}}\,{\cal{H}}\,{\cal{U}}^{-1}\,|\Psi_0\rangle$
with $|\Psi_0\rangle\equiv{\cal U}|\Phi_0\rangle$. and the
transformed hamiltonian
$\widetilde{{\cal{H}}}\equiv{\cal{U}}\,{\cal{H}}\,{\cal U}^{-1}$
defined by Eq.~(\ref{h.tilde}); if $|\Phi_0\rangle$ belongs to the
reduced Hilbert space, it is
\begin{equation}
|\Psi_0\rangle\equiv {\cal U}|\Phi_0\rangle=
\sum_{\{s \}} c_{\{s\}}
|s_1 s_2\,\cdots\, s_N\rangle
\label{e.trialgs}
\end{equation}
with $\{s\}\,{\equiv}\,(s_1,s_2,s_3...s_N)$, and
$s_i\,{\equiv}\,\langle\Psi_0|S_i^z|\Psi_0\rangle\,{=}\,{+}1,0$.

The simplest trial ground state allowing the description of soliton
pairs dissociation is that defined by Eq.~(\ref{e.trialgs}) with
$c_{\{s\}}\,{=}\,t_{s_1s_2s_3}t_{s_2s_3s_4}\,
\cdots\,t_{s_{N-2}s_{N-1}s_N}$\,.
The variational parameters are the six amplitudes $t_{+++}$,
$t_{++0}\,{=}\,t_{0++}$, $t_{+0+}$, $t_{0+0}$,
$t_{00+}\,{=}\,t_{+00}$, $t_{000}$, where
$|t_{s_{i-1}s_{i}s_{i+1}}|^2$ represents the probability for
$(S^z_{i-1},S^z_{i},S^z_{i+1})$ to be equal
$(s_{i-1},s_{i},s_{i+1})$; a common arbitrary factor may be used for
normalizing $|\Psi_0 \rangle$. We notice that the chosen form for
$c_{\{s\}}$ is such that the probability for $|\Psi_0\rangle$ to
contain coupled solitons is finite independently of that relative to
the occurrence of isolated solitons, whose presence is unambiguously
marked by $t_{+0+}\,{\neq}\,0$.

Without going into the details of the variational calculations,
reported in Ref.~\onlinecite{MikeskaV1995}, we here discuss our final
results. Due to the normalization condition, the number of
variational parameters is reduced from six to five; moreover, the
energy $\langle\Psi_0|{\cal U}{\widetilde{\cal H}} {\cal
U}^{-1}|\Psi_0\rangle$ is found to depend just on four precise
combinations of the original parameters,
\begin{eqnarray}
 \gamma\equiv|t_{++0}|^2|t_{00+}|^2 &\,\,\,\Rightarrow\,\,\,
  & (\cdots\,++00++\,\cdots)\nonumber\\
 \pi\equiv|t_{++0}|^2|t_{+0+}| &\,\,\,\Rightarrow\,\,\,
  & (\cdots\,++0++\,\cdots)\nonumber\\
 \chi\equiv |t_{00+}|^2|t_{0+0}| &\,\,\,\Rightarrow\,\,\,
  & (\cdots\,00+00\,\cdots)\nonumber\\
\rho\equiv|t_{000}|&\,\,\,\Rightarrow\,\,\,& (\cdots\,000\,\cdots)~,
\label{varpar}
\end{eqnarray}
whose square moduli are related to the probabilities that the
corresponding strings ($\,\,\Rightarrow\,\,$) be contained in
$|\Psi_0\rangle$; in particular, $\gamma^2$ and $\pi^2$ refer to the
probabilities for coupled and isolated solitons, respectively, to
occur in the ground state.

\begin{figure}
\includegraphics[height=85mm,angle=90]{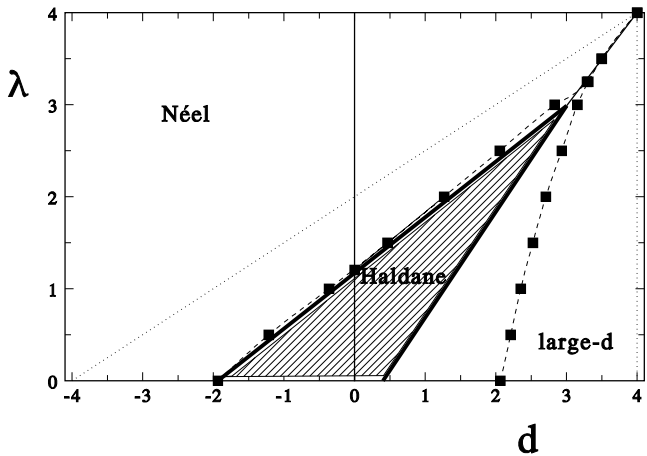}
\caption{\label{f.hal1}
Phase diagram in the $\lambda\,{>}\,0$ half-plane: our results
(squares) are shown together with those of
Ref.~\onlinecite{KennedyT1992} (dotted lines); the Haldane phase
should correspond to the shaded area, according to the best available
numerical data~\cite{BotetJK1983} (solid lines).}
\end{figure}

Both the analytical expression for the energy and the numerical
minimization show that it exists a critical value
$\lambda_{\rm{c}}\,{=}\,\lambda_{\rm{c}}(d)\,{>}\,d$ such that, for
$\lambda\,{>}\,\lambda_{\rm{c}}$ the minimal energy is attained for
$\pi\,{=}\,\rho\,{=}\,0$; the condition
$\lambda\,{=}\,\lambda_{\rm{c}}(d)$ can hence define a curve of phase
separation. We therefore single out three different phases,
characterized by
\begin{equation}
(a)~\pi\,{=}\,\rho\,{=}\,0,~~~(b)~{\rm{all~parameters}}\neq
0,~~(c)~\chi\,{=}\,t_{+++}\,{=}\,0
\label{e.abc}
\end{equation}
in the ground state. The corresponding phase diagram is shown in
Fig.~\ref{f.hal1}, together with that obtained with a factorized
trial ground state~\cite{KennedyT1992}, and by numerical
simulations~\cite{BotetJK1983}.

The $(a){-}(b)$ transition is seen to quite precisely describe the
N\'eel$-$Haldane one, and this leads us to define the condition
$(a)$, meaning the occurrence of exclusively coupled solitons, as
typical of the N\'eel phase. As for the $(c){-}(b)$ transition, it is
to be noticed that the use of the reduced Hilbert space is not fully
justified in the $\lambda\,{<}\,d$ region, where we in fact do not
expect quantitatively precise results.

As for a comparison between our results and the exact numerical data
available, we have considered, along the $d\,{=}\,0$ axis, two
specific quantities: the critical anisotropy $\lambda_c(d)$, where
the N\'eel phase becomes unstable with respect to the Haldane one,
and the ground-state energy $E_0(d,\lambda)$ at the isotropic point
$\lambda\,{=}\,1$. For the critical anisotropy we find
$\lambda_c(0)\,{=}\,1.2044(5)$ to be compared with the value obtained
by exact diagonalization~\cite{SakaiT1990}, $\lambda_c(0)
\approx 1.19$; for the energy we find $E_0(0,1)\,{=}\,-1.3663(5)$
to be compared with $E_0(0,1)\,{=}\, -1.4014(5)$, again from exact
diagonalization technique~\cite{GolinelliJL1994}; the value obtained
with the factorized trial ground state is~\cite{KennedyT1992}
$E_0(0,1)\,{=}\,-4/3$.

\begin{figure}
\includegraphics[height=85mm,angle=90]{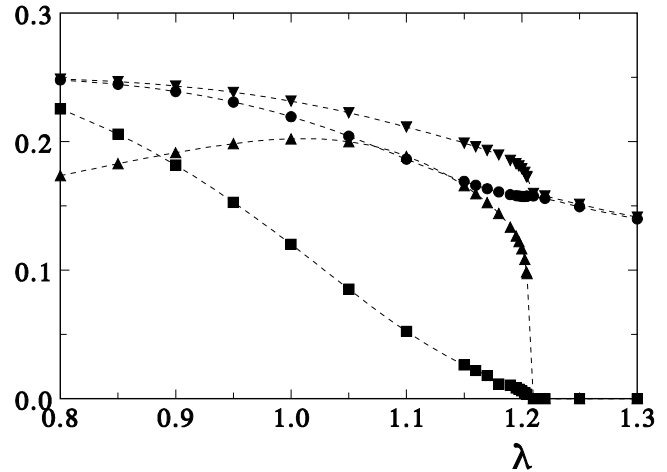}
\caption{\label{f.hal2}
Parameters $w_{(1)}$ (squares), $w_{(2)}$ (circles), $w_{(3)}$
(upward triangles), and $w_{(2,2)}$ (downward triangles), for
$d\,{=}\,0$.}
\end{figure}

In Fig.~\ref{f.hal2} we show the variational parameters as $\lambda$
is varied with $d\,{=}\,0$, i.e. along the $y$ axis of the
phase-diagram; in fact, rather than the parameters with respect to
which we have actually minimized the energy, the following
combinations are considered:
\begin{eqnarray}
w_{(1)}=\pi^2/t^2_{+++}&;&
w_{(2)}=\gamma~; \nonumber\\
w_{(2,2)}=(\chi~\gamma~t_{+++})^{1/2}&;& w_{(3)}=t_{+++}^2
(\rho~\gamma)^{2/3}~.
\label{wp}
\end{eqnarray}

The above quantities have a straightforward physical meaning, as they
are directly related with the probabilities for a soliton to appear
along the chain as an isolated excitation $(w_{(1)})$, as part of a
soliton pair $(2 w_{(2)}^2)$, as part of a string made of three
adjacent solitons $(3 w_{(3)}^3)$, and finally as part of a string
made of two soliton pairs separated by one site $(4 w_{(2,2)}^4)$.
From Fig.~\ref{f.hal2} it turns evident that the Haldane phase is
featured by the occurrence of isolated solitons ($w_{(1)} \ne 0$), as
well as of strings made of three adjacent solitons ($w_{(3)}{\ne}0$).

This result confirms that, as elicited by the analysis of the
phase-diagram, the Haldane phase is characterized by our condition
$(b)$.

Given their essential role, we have also studied the $x$ and $z$
component of the string order parameter, as well as the solitons
density $n_0\,{=}\,1-\big\langle(S^z)^2\big\rangle$. After KT we
expect $O_{\rm{string}}^z({\cal H})\neq 0$ in both the N\'eel and the
Haldane phase, and $O_{\rm{string}}^x({\cal H})\neq 0$ just in the
Haldane phase. In fact, analytical expressions for $O^x$ and $O^z$
may be written~\cite{MikeskaV1995} in terms of four of the five
variational parameters (\ref{wp}), and show that

\noindent - $O^x_{\rm{string}}({\cal H})\,{=}\,0$ if $\pi\,{=}\,\rho\,{=}\,0$ or
$\chi\,{=}\,t_{+++}\,{=}\,0$, i.e. in phase ($a$) and $(c)$;

\noindent - $O^z_{\rm{string}}({\cal H})\,{>}\,0$ in all phases, asymptotically
vanishing as $\rho\,{\to}\,1$, i.e. in the far large-$d$ phase.

\begin{figure}
\includegraphics[height=85mm,angle=90]{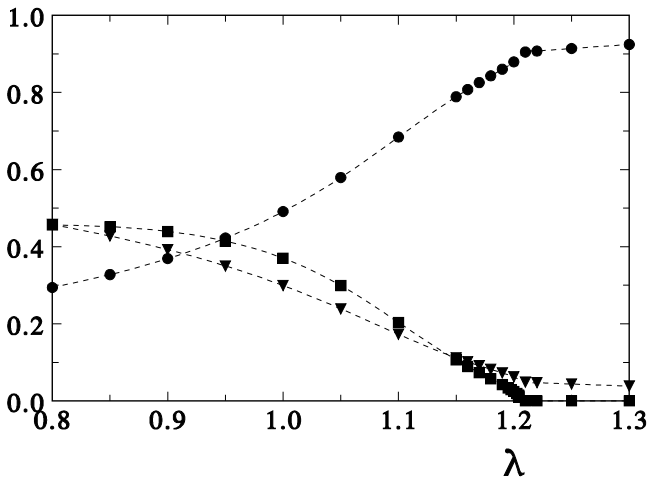}
\caption{\label{f.hal3}
String order parameters $O^x_{\rm{string}}$ (squares),
$O^z_{\rm{string}}$ (circles), and solitons density $n_0$
(triangles), for $d\,{=}\,0$.}
\end{figure}

\begin{figure}
\includegraphics[height=85mm,angle=90]{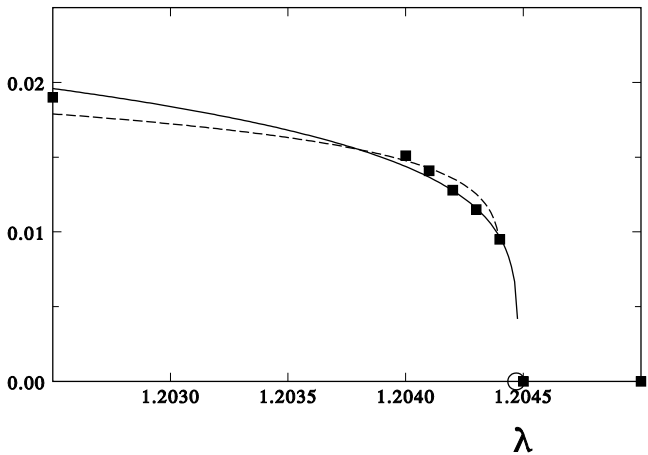}
\caption{\label{f.hal4}
Critical behavior of $O^x_{\rm{string}}$ for $d\,{=}\,0$: squares are
results; curves are obtained by best-fit procedure from
$O^x_{\rm{string}}\,{\sim}\,(\lambda-\lambda_c)^\beta$ with $\beta$
fixed to $0.125$ (dashed curve), and as fitting parameter, resulting
in $\beta\,{=}\,0.217$ (solid curve). Both procedures give
$\lambda_{\rm{c}}\,{=}\,1.2044(5)$, marked by a circle in figure.}
\end{figure}

In more details, we notice that $O^x_{\rm{string}}\,{=}\,0$ whenever
the ground state does not contain strings made of an odd number of
adjacent spins; as soon as the shortest string of such type, namely
the isolated soliton, appears along the chain, then
$O^x_{\rm{string}}$ gets finite. The unphysical result
$O^z_{\rm{string}}\,{>}\,0$ in the $(c)$ phase, vanishing only as
$d\,{\to}\,\infty$ rather than everywhere in the large-$d$ phase, is
due to our assuming the ground state to belong to the reduced Hilbert
space, which is actually licit just in the $\lambda\,{>}\,d$ region.

In Fig.~\ref{f.hal3} we show $O^x_{\rm{string}}$,
$O^z_{\rm{string}}$, and $n_0$ as $\lambda$ varies with $d\,{=}\,0$:
We underline that $O^x_{\rm{string}}$ gets finite continuously but
with discontinuous derivative at the transition (reflecting the
behavior of $w_{(1)}$ and $w_{(3)}$ shown in Fig.~\ref{f.hal2}), so
that the N\'eel$-$Haldane quantum phase transition is recognized as a
second order one. In Fig.~\ref{f.hal4} we zoom the order parameter
$O^x_{\rm{string}}$ around the critical point: its behavior is seen
to be described by a power law $O^x_{\rm{string}}\,\sim\,(\lambda_c -
\lambda)^{\beta}$, as expected for a continuous phase transition;
our estimated value for the critical exponent is
$\beta\,{=}\,0.217(5)$ to be compared with $\beta\,{=}\,0.125$,
corresponding to the Ising model in a transverse field, to whose
universality class the Haldane transition is suggested to belong
to~\cite{GomezSantos1989}. At the isotropic point
($d\,{=}\,0,~\lambda\,{=}\,1$) we find
$O^x_{\rm{string}}\,{=}\,0.3700(5)$ in full agreement with the value
obtained by exact diagonalization~\cite{ElstnerM1994}.

The overall good agreement between our results and the numerical
available data, allows us to conclude that the N\'eel-Haldane
transition is a second-order one, and that the string order parameter
$O^x_{\rm{string}}$, revealing {\em hidden} order along the $x$
direction, is the appropriate order parameter for the Haldane phase.
The disordered ground state featuring the Haldane phase is seen to
originate by soliton pairs dissociation, according to this path:
Solitons occur just in pairs in the antiferromagnetically ordered
N\'eel phase; at the N\'eel-Haldane transition soliton pairs
dissociate and the byproducts rearrange in strings made of an odd
number of solitons. These strings are ultimately responsible for the
disorder of the ground state.

\section{Two-dimensional isotropic Heisenberg model}
\label{s.isotropic}

The 2D isotropic QHAF on the square lattice is one of the magnetic
models most intensively investigated in the last two decades. This is
due both to its theoretically challenging properties and to its being
considered the best candidate for modeling the magnetic behavior of
the parent compounds of some high-$T_{\rm{c}}$
superconductors~\cite{SokolP1993,Chakravarty1990}.

From a theoretical point of view the fully isotropic Heisenberg model
in $d$ dimensions, thanks to the simple structure of its Hamiltonian
(whose high symmetry is responsible for most of its peculiar
features), may well be considered a cornerstone of the modern theory
of critical phenomena, with its relevance extending well beyond the
only magnetic systems. The $d\,{=}\,2$ case earned additional
interest, representing the boundary dimension separating systems with
and without long-range order at finite
temperature~\cite{MerminW1966}. The antiferromagnetic coupling adds
further appeal, as the classical-like N\'eel state is made unstable
by quantum fluctuations and the ground state of the system is not
exactly known. It can be rigorously proven~\cite{NevesP1986} to be
ordered for $S{\ge}1$; for $S\,{=}\,1/2$ there is no rigorous proof,
although evidences for an ordered ground state can be drawn from many
different studies (for a review, see, for instance,
Ref.~\onlinecite{Manousakis1991}).

On the experimental side the attention on the properties of 2D QHAF
was mainly triggered by the fact that among the best experimental
realizations of this model we find several parent compounds of
high-$T_{\rm{c}}$ superconductors, as, e.g., La$_2$CuO$_4$ or
Sr$_2$CuO$_2$Cl$_2$~\cite{KastnerBSE1998,GrevenBEKKMST1994,GrevenBEKMS1995},
both having spin $S\,{=}\,1/2$. In such materials, as well as in
other magnetic compounds with a layered crystal structure as
La$_2$NiO$_4$~\cite{NakajimaYHEGB1995} and
K$_2$NiF$_4$~\cite{GrevenBEKKMST1994,GrevenBEKMS1995} ($S\,{=}\,1$),
Rb$_2$MnF$_4$~\cite{LeeGWBS1998} and KFeF$_4$~\cite{FultonEtal94}
($S\,{=}\,5/2$) or copper formate tetradeuterate (CFTD,
$S\,{=}\,1/2$)~\cite{CCCMRTV2000prl} the magnetic ions form parallel
planes and interact strongly only if belonging to the same plane. The
interplane interaction in these compounds is several orders of
magnitude smaller than the intraplane one, thus offering a large
temperature region where their magnetic behavior is indeed 2D down to
those low temperatures where the weak interplane interaction becomes
relevant, driving the system towards a 3D ordered phase: an
antiferromagnetic Heisenberg interaction and the small spin value
make these compounds behave as 2D QHAFs. Even the onset of 3D
magnetic long-range order is however strongly affected by the 2D
properties of the system: indeed, the observed 3D magnetic transition
temperature is comparable with the intraplane interaction energy,
i.e., several order of magnitude larger than that one can expect only
on the basis of the value of the interplane coupling. Such apparently
odd behavior can be easily understood by observing that the
establishing of in-plane correlations on a characteristic distance
$\xi$ effectively enhances the interplane coupling by a factor
$(\xi/a)^2$, $a$ being the lattice constant. The latter consideration
is one of the reasons explaining why most of the attention, both from
the experimental and theoretical point of view, was devoted to the
low-temperature behavior of the correlation length $\xi$ of the 2D
QHAF (in the following $\xi$ will be always given in units of the
lattice constant $a$).

The 2D QHAF is described by the Hamiltonian
\begin{equation}
 {\cal H}= \frac J2 \sum_{{\bm{i}},{\bm{d}}}
 {\bm{S}}_{\bm{i}}{\cdot}
 {\bm{S}}_{{\bm{i}}+{\bm{d}}}~,
\label{e.ham}
\end{equation}
where $J$ is positive and the quantum spin operators
${\bm{S}}_{\bm{i}}$ satisfy $|{\bm{S}}_{\bm{i}}|^2\,{=}\,S(S{+}1)$.
The index ${\bm{i}}\equiv(i_1,i_2)$ runs over the sites of a square
lattice, and ${\bm{d}}$ represents the displacements of the 4
nearest-neighbors of each site, $(\pm{1},0)$ and $(0,\pm{1})$.

In addition to the first approximations usually employed to
investigate the low temperature properties of magnetic systems as,
e.g., mean-field and (modified) spin-wave theory, the critical
behavior of the 2D QHAF was commonly interpreted on the basis of the
results obtained by field theory starting from the so-called 2D {\em
quantum nonlinear $\sigma$ model}
(QNL$\sigma$M)~\cite{ChakravartyHN1989}, whose action is given by
\begin{equation}
 S=\frac 1{2g}\,\int d{\bm x}\int_0^{u}\,d\tau\,
 \big(|\nabla{\bm n}|^2+|\partial_\tau{\bm n}|^2\big)\,;
 ~~|{\bm n}|^2=1\,.\label{NLSM}
\end{equation}
In the last Equation ${\bm{n}}({\bm{x}})$ is a unitary 3D vector
field, $g\,{=}\,c\Lambda/\rho$ and $u\,{=}\,c\Lambda/T$ are the
coupling and the imaginary-time cut-off respectively, and the two
parameters $\rho$ and $c$ are usually referred to as {\em spin
stiffness} and {\em spin-wave velocity}. Despite their names, the two
parameters $\rho$ and $c$ are however just phenomenological fitting
constants which can be rigorously related to the proper parameter $J$
and $S$ of the original magnetic Hamiltonian (\ref{e.ham}) only in
the large-$S$ limit~\cite{Haldane1983,Affleck1985}. The source of
non-linearity in the model Eq.~(\ref{NLSM}), which is seemingly
quadratic in the field variables, is the constraint imposed onto the
length of the field $\bm{n}$.

The relation between the 2D QHAF and the QNL$\sigma$M was first
exploited to interpret the experimental data on cuprous oxides by
Chakravarty, Halperin, and Nelson~\cite{ChakravartyHN1989} (CHN) who
used symmetry arguments to show that the long-wavelength physics of
the QHAF is the same of that of the QNL$\sigma$M; in other words the
physical observables of the two models show the same functional
dependence upon $T$, if the long-wavelength excitations are assumed
to be the only relevant ones, as one expects to be at low
temperature.

The analysis carried out by CHN on the QNL$\sigma$M lead to single
out three different regimes, called {\em quantum disordered}, {\em
quantum critical} (QCR) and {\em renormalized classical} (RCR), the
most striking difference amongst them being the temperature
dependence of the spin correlations. If $g$ is such as to guarantee
LRO at $T\,{=}\,0$, the QNL$\sigma$M is in the RCR at very
low-temperature and the correlation length $\xi$ behaves
as~\cite{HasenfratzN1991}:
\begin{equation}
 \xi_{\rm{3l}}= \frac e8 \Big(\frac c{2\pi\rho}\Big)
 \exp \Big( \frac{2\pi\rho} T\Big)~
 \Big[1-\frac T{4\pi\rho}\Big]~.
\label{e.xiCH2N2}
\end{equation}
CHN found also that by raising the temperature any 2D QNL$\sigma$M
with an ordered ground state crosses over from the RCR to the QCR,
characterized by a correlation length
$\xi\propto\alpha(T)\,{=}\,c/T$.

The first direct comparison between experimental data on spin $1/2$
compounds and the prediction of the QNL$\sigma$M field theory in the
RCR gave surprisingly good agreement and caused an intense activity,
both theoretical and experimental, in the subsequent years. However,
with the accumulation of new experimental data on higher spin
compounds it clearly emerged that the experimentally observed
behavior of $\xi(T)$ for larger spin could not be reproduced neither
by the original simplified (2-loop) form of Eq.~(\ref{e.xiCH2N2})
given by CHN (which does not contain the term in square brackets),
nor by the three-loops result (\ref{e.xiCH2N2}) derived by Hasenfratz
and Niedermayer~\cite{HasenfratzN1991} (HN); moreover no trace of QCR
behavior was found in pure compounds. The discrepancies observed
could be due to the fact that the real compounds do not behave like
2D QHAF or to an actual inadequacy of the theory. In particular the
CHN-HN scheme introduces two possible reasons for such inadequacy to
occur: the physics of the 2D QHAF is not properly described by that
of the 2D QNL$\sigma$M and/or the two(three)-loop
renormalization-group expressions derived by CHN-HN do hold at
temperatures lower than those experimentally accessible. After an
almost ten years long debate, the latter possibility has finally
emerged as the correct one, being strongly supported not only by our
own work, but also by other independent theoretical
approaches~\cite{Elstner97Etal95}, joined with the analysis of the
experimental and the most recent quantum Monte Carlo (QMC) data for
the 2D QHAF.

The theoretical approach we employed to investigate the 2D QHAF is
the effective Hamiltonian
method~\cite{CTVV1996prl,CTVV1997prb,CTVV1998prb}, developed within
the framework of the {\em pure-quantum self-consistent harmonic
approximation} (PQSCHA) we introduced at the beginning of the
90's~\cite{CTVV1992pra,CGTVV1995jpcm}. The PQSCHA starts from the
Hamiltonian path-integral formulation of statistical mechanics which
allows one to separate in a natural way classical and quantum
fluctuations: only the latter are then treated in a self-consistent
harmonic approximation, finally getting an effective classical
Hamiltonian, whose properties can thereafter be investigated by all
the techniques available for classical systems. The idea of
separating classical and quantum fluctuations turned out to be
fruitful not only in view of the implementation of the PQSCHA, but
also in the final understanding of the connection between
semiclassical approaches and quantum field
theories~\cite{BCVV2003prb}, which could be possible also thanks to
the paper by Hasenfratz~\cite{Hasenfratz2000} about corrections to
the field-theoretical results due to cutoff effects.

The PQSCHA naturally applies to bosonic systems, whose Hamiltonian is
written in terms of conjugate operators $\underline{\hat{q}}\equiv
(\hat{q}_1,...\hat{q}_N)$, $\underline{\hat{p}}\equiv
(\hat{p}_1,...\hat{p}_N)$ such that
$[\hat{q}_{m},\hat{p}_{n}]\,{=}\,i\delta_{mn}$; the method, however,
does not require ${\cal H}(\underline{\hat{p}},\underline{\hat{q}})$
to be standard, i.e., with separate quadratic kinetic
$\underline{\hat{p}}$-dependent and potential
$\underline{\hat{q}}$-dependent terms, and its application may be
extended also to magnetic systems, according to the following
scheme~\cite{CGTVV1995jpcm}: The spin Hamiltonian
${\cal{H}}(\underline{\bm{S}})$ is mapped to
${\cal{H}}(\underline{\hat{p}},\underline{\hat{q}})$ by a suitable
spin-boson transformation; once the corresponding Weyl symbol ${\cal
H}(\underline{p},\underline{q})$, with $\underline{p}\equiv
(p_1,...p_N)$ and $\underline{q}\equiv (q_1,...q_N)$ classical
phase-space variables, has been determined, the PQSCHA
renormalizations may be evaluated and the effective classical
Hamiltonian ${\cal{H}}_{\rm{eff}}(\underline{p},\underline{q})$ and
effective classical function
${\cal{O}}_{\rm{eff}}(\underline{p},\underline{q})$ corresponding to
the observable ${\cal O}$ of interest follow. Finally the effective
functions ${\cal{H}}_{\rm{eff}}(\underline{\bm{s}})$ and
${\cal{O}}_{\rm{eff}}(\underline{\bm{s}})$, both depending on
classical spin variables ${\bm{s}}$ with $|{\bm{s}}|\,{=}\,1$ and
containing temperature- and spin-dependent quantum renormalized
parameters, are reconstructed by the inverse of the classical
analogue of the spin-boson transformation used at the beginning.

In order to successfully carry out such renormalization scheme, the
Weyl symbol of the bosonic Hamiltonian must be a well-behaved
function in the whole phase space. Spin-boson transformations, on the
other hand, can introduce singularities as a consequence of the
topological impossibility of a global mapping of a spherical phase
space into a flat one. The choice of the transformation must then be
such that the singularities occur for configurations which are not
thermodynamically relevant, and whose contribution may be hence
approximated. Most of the methods for studying magnetic systems do in
fact share this problem with the PQSCHA; what makes the difference is
that by using the PQSCHA one separates the classical from the
pure-quantum contribution to the thermal fluctuations, and the
approximation only regards the latter, being the former exactly taken
into account when the effective Hamiltonian is recast in the form of
a classical spin Hamiltonian.

The spin-boson transformation which constitutes the first step of the
magnetic PQSCHA is chosen according to the symmetry properties of the
original Hamiltonian and of its ground state. In the case of the 2D
QHAF both Dyson-Maleev and Holstein-Primakoff transformation can be
employed finally obtaining~\cite{CTVV1997prb}:
\begin{eqnarray}
 \frac{{\cal H}_{\rm{eff}}}{J\widetilde{S}^2}
 &=& \frac{\theta^4}2 \sum_{{\bm{i}},{\bm{d}}} {\bm{s}}_{\bm{i}}
 {\cdot} {\bm{s}}_{{\bm{i}}+{\bm{d}}} + N\,{\cal G}(t)~,
\label{e.Heff}
\\
 {\cal{G}}(t) &=& \frac tN\sum_{\bm{k}}\ln\frac{\sinh{f_k}}{\theta^2f_k}
 - 2 \kappa^2 {\cal{D}}~,
\end{eqnarray}
with the temperature and spin dependent parameters
\begin{eqnarray}
 \theta^2&=&1-\frac{{\cal D}}2~,
\label{e.theta}
\\
 {\cal D}&=&\frac1{\widetilde S\, N} \sum_{\bm{k}}
 (1-\gamma_{\bm{k}}^2)^{\frac12}{\cal L}_{\bm{k}}~,
\label{e.D}
\\
 f_{\bm{k}} &=& \frac{\omega_{\bm{k}}}{2\widetilde{S}t}~,
 ~~~ {\cal L}_{\bm{k}} = \coth f_{\bm{k}}-\frac1{f_{\bm{k}}}~.
\label{e.fk}
\end{eqnarray}
In the previous Equations
$\gamma_{\bm{k}}\,{=}\,(\cos{k_1}\,{+}\,\cos{k_2})/2$, $N$ is the
number of sites of the lattice and ${\bm{k}}\equiv(k_1,k_2)$ is the
wave-vector in the first Brillouin zone;
$\widetilde{S}\,{\equiv}\,S\,{+}\,1/2$ is the effective classical
spin length, which naturally follows from the renormalization scheme,
and $t\,{\equiv}\,T/J\widetilde{S}^2$ is the reduced temperature
defined in terms of the energy scale $J\widetilde{S}^2$. The
renormalization scheme is closed by the self-consistent solution of
the two coupled equations
$\omega_{\bm{k}}\,{=}\,4\kappa^2(1-\gamma^2_{\bm{k}})^{1/2}$ and
$\kappa^2\,{=}\,\theta^2-t/(2\kappa^2)$\,. The {\em pure-quantum}
renormalization coefficient ${\cal{D}}\,{=}\,{\cal{D}}(S,t)$ takes
the main contribution from the high-frequency part (short-wavelength)
of the spin-wave spectrum, because of the appearance of the Langevin
function ${\cal L}_{\bm{k}}$. ${\cal{D}}$ measures the strength of
the pure-quantum fluctuations, whose contribution to the
thermodynamics of the system is the only approximated one in the
PQSCHA scheme. The theory is hence quantitatively meaningful as far
as ${\cal{D}}$ is small enough to justify the self-consistent
harmonic treatment of the pure-quantum effects. In particular, the
simple criterion ${\cal{D}}\,{<}\,0.5$ is a reasonable one to assess
the validity of the final results.

The most relevant information we get from Eq. (\ref{e.Heff}) is that
the symmetry of the Hamiltonian is left unchanged so that from a
macroscopic point of view the quantum system essentially behaves, at
an actual temperature $t$, as its classical counterpart does at an
effective temperature $t_{\rm{eff}}\,{=}\,t/\theta^4(S,t)~$. This
allows us to deduce the behavior of many observables (but not all\,!,
see Refs.~\onlinecite{CTVV1996prl,CTVV1997prb, CTVV1998prb} for
details) directly from the behavior of the corresponding classical
quantities. This is the case of the correlation length, which turns
out to be given simply by:
\begin{equation}
\xi(t)=\xi_{\rm{cl}}(t_{\rm{eff}})
\label{e.xi}
\end{equation}
so that once $\theta^4(S,t)~$ has been evaluated, the only additional
information we need is the classical $\xi_{\rm{cl}}(t)$, which is
available from {\em classical} Monte Carlo simulation and analytical
asymptotic expressions~\cite{CFT} as $t\,{\to}\,0$.

\begin{figure}
\includegraphics[width=85mm,angle=0]{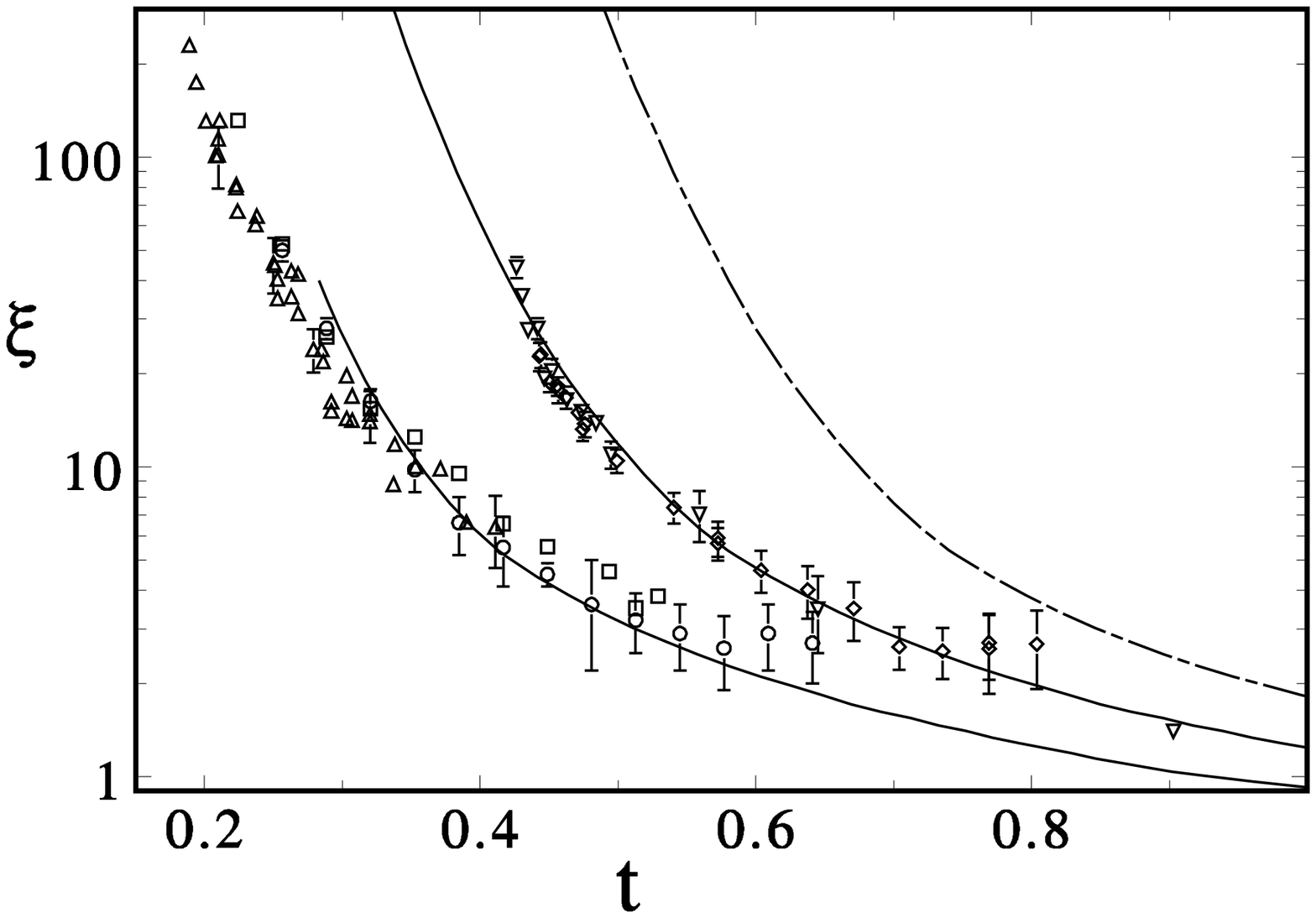}
\caption{
Correlation length $\xi$ vs $t$, for $S\,{=}\,1/2$ (leftmost) and
$S\,{=}\,1$. The symbols are experimental data; for $S\,{=}\,1/2$:
$^{63}$Cu~NQR data~\cite{CarrettaRS1997} (circles) and neutron
scattering data for La$_2$CuO$_4$ (squares~\cite{KastnerBSE1998}) and
for Sr$_2$CuO$_2$Cl$_2$
(up-triangles~\cite{GrevenBEKKMST1994,GrevenBEKMS1995}); for
$S\,{=}\,1$: neutron scattering data for La$_2$NiO$_4$
(down-triangles~\cite{NakajimaYHEGB1995}) and for K$_2$NiF$_4$
(diamonds~\cite{GrevenBEKKMST1994,GrevenBEKMS1995}). The classical
result (dash-dotted line) is also reported.
\label{f.xiS05-1}
}
\end{figure}

\begin{figure}
\includegraphics[width=85mm,angle=0]{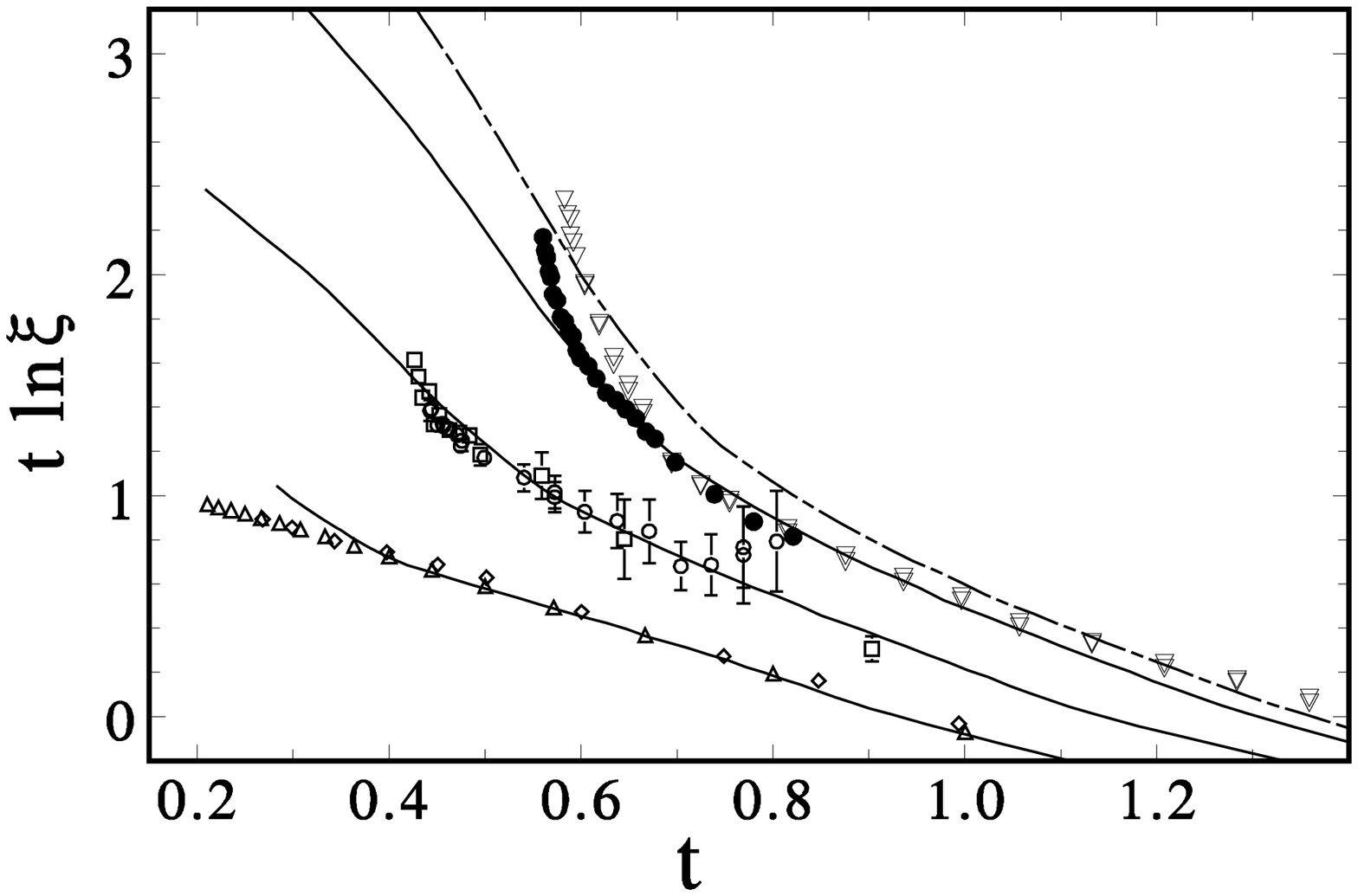}
\caption{
The function $y(t)\,{=}\,t\ln\xi$ vs $t$, for (from the rightmost
curve) $S\,{=}\,\infty$, $5/2$, $1$ and $1/2$; the
up-triangles~\cite{KimLT1997} and the diamonds~\cite{MakivicD1991}
are quantum MC data for $S\,{=}\,1/2$; also reported are neutron
scattering data for La$_2$NiO$_4$ (open
circles~\cite{NakajimaYHEGB1995}), for K$_2$NiF$_4$
(squares~\cite{GrevenBEKKMST1994,GrevenBEKMS1995}), KFeF$_4$ (filled
circles~\cite{FultonEtal94}) and Rb$_2$MnF$_4$
(down-triangles~\cite{LeeGWBS1998}). The abrupt rising of the
experimental data for the $S\,{=}\,5/2$ compounds at
$t\,{\simeq}\,0.65$ is due to the effect of the small, but finite,
anisotropies and will be discussed in more details in
Section~\ref{s.anisotropic}.
\label{f.tlnxi}
}
\end{figure}

Sample results obtained by PQSCHA are shown in the figures. In
Fig.~\ref{f.xiS05-1} the correlation length for $S\,{=}\,1/2$ and
$S\,{=}\,1$ is compared with experimental data; a similar comparison,
including MC data for $S\,{=}\,1/2$ and experimental data on
$S\,{=}\,5/2$ compounds KFeF$_4$ and Rb$_2$MnF$_4$ is made in
Fig.~\ref{f.tlnxi}, but along the vertical axis the quantity
$t\ln\xi$ is reported in order to better appreciate the deviation
from the predicted RCR behavior, that would correspond to a straight
line at low $t$. From the last picture one can easily see that both
PQSCHA curves and experimental data for $S\ge 1$ (including the exact
$S\,{=}\,\infty$ classical result) display a change of slope at
intermediate temperature, followed by a curvature inversion at lower
$t$. On the other hand, by looking at the $S\,{=}\,1/2$ case it
becomes clear why the QNL$\sigma$M approach gave such a good
agreement when firstly used to fit the experimental data. The change
in both the slope and the curvature of $t\ln\xi$ is less pronounced
and possibly occurs at lower temperatures, the lower the spin: in the
$S\,{=}\,1/2$ case, it is difficult to say whether these features are
still present or not, but, if yes, they occur in a temperature region
where the extremely high value of $\xi$ ($\approx 10^4$) makes both
the experimental and the simulation data more difficult to be
obtained.

After having realized that the field theoretical prediction by CHN
could not be applied to the $S\ge1$ 2D QHAF in the temperature range
probed by the experiments, the following questions were waiting for a
satisfactory answer: $(i)$ the real range of applicability of the
asymptotic three-loop expression (\ref{e.xiCH2N2}) at different $S$,
and $(ii)$ the possible extension of the PQSCHA results to lower
temperature, both in view of $(iii)$ a comprehensive description of
the behavior of the correlation length of the 2D QHAF in the entire
range of temperature and spin values.

\begin{figure}
\includegraphics[width=85mm,angle=0]{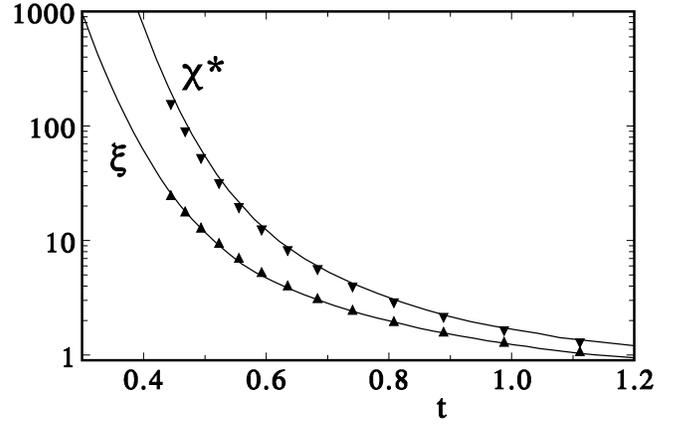}
\caption{
Correlation length $\xi$ and staggered susceptibility
$\chi^*\,{\equiv}\,\chi/\widetilde{S}^2$ vs $t$ for $S\,{=}\,1$.
Symbols are QMC data from Ref.~\onlinecite{HaradaTK1998}.
\label{f.xichis10}
}
\end{figure}

A substantial contribution to settle this conundrum came only from
QMC simulations for higher spin values able to probe the very large
correlation length region~\cite{BeardEtal1998-2000}: indeed, high
precision Monte Carlo data for $S\,{=}\,1$ and moderate correlation
length could still be very well interpreted by PQSCHA and did not
display the RCR asymptotic behavior, as shown~\cite{CTVV1998prb} in
Fig.~\ref{f.xichis10}, where we compared our curves for $\xi$ and
staggered susceptibility $\chi^*$ with QMC data obtained by Harada
et~al.~\cite{HaradaTK1998}.

QMC results for $\xi$ by Beard and
coworkers~\cite{BeardEtal1998-2000} showed unambiguously that the
three-loop Eq.~(\ref{e.xiCH2N2}) holds only for temperatures low
enough to ensure an extremely large correlation length, e.g.,
$\xi\gtrsim{10^5}$ for $S\,{=}\,1$, $\xi\gtrsim{10^{12}}$ for
$S\,{=}\,3/2$, and generally cosmological correlation lengths for
$S\,{>}\,3/2$, thus definitely excluding any possibility of employing
QNL$\sigma$M results to interpret available experimental data.

In Ref.~\onlinecite{Hasenfratz2000} Hasenfratz showed why cutoff
effects, which are so devious for $S\,{=}\,1/2$, significantly modify
the correlation length for $S\ge{1}$. Reaching such goal was possible
only going back to a direct mapping between the QHAF and the
QNL$\sigma$M, and the resulting cutoff-corrected field-theoretical
outcome is~\cite{Hasenfratz2000}
\begin{equation}
 \xi_{{}_{\rm{H}}}(T,S) = \xi_{\rm{3l}}(T,S)~e^{-C(T,S)}~,
\label{e.xiH}
\end{equation}
where $C(T,S)$, defined in Eq.~(14) of
Ref.~\onlinecite{Hasenfratz2000}, is an integral of familiar
spin-wave quantities over the first Brillouin zone.

With this correction, which is the leading order in the spin-wave
expansion for the cutoff correction, it is possible to obtain
numerically accurate agreement with QMC data down to $\xi\gtrsim
10^3$ for all $S$.

\begin{figure}
\includegraphics[width=85mm,angle=0]{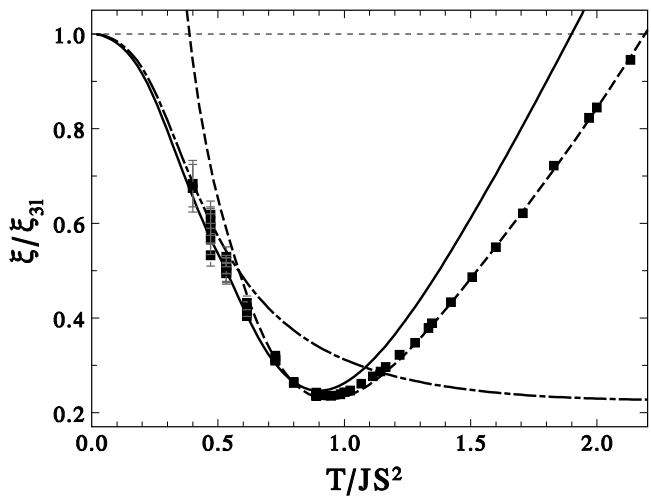}
\caption{\label{f.Fig23beard}
Ratio $\xi/\xi_{\rm{3l}}$ {\em vs} $T/JS^2$ for $S\,{=}\,5/2$. Solid
line: $\xi_{\rm{cl}}(t_{\rm{eff}}^{\rm{H}})$; dash-dotted line:
$\xi_{\rm{H}}\,{\simeq}\,\xi_{\rm{3l}}^{\rm{cl}}(t_{\rm{eff}}^{\rm{H}})$;
dashed line: standard PQSCHA result; symbols are QMC
data~\cite{BCVV2003prb}. }
\end{figure}

In our most recent paper~\cite{BCVV2003prb} about 2D QHAF we showed
that by employing the explicit expression for $C(T,S)$, and by
substituting in Eq. (\ref{e.xiCH2N2}) the spin stiffness $\rho$ and
the spin-wave velocity $c$ given by the mapping of QHAF onto the
QNL$\sigma$M, the leading terms of the result (\ref{e.xiH}) not only
can be cast into the form
$\xi_{{}_{\rm{H}}}(T,S)\,{=}\,\xi_{\rm{3l}}^{\rm{cl}}(t_{\rm{eff}}^{\rm{H}})$,
in strict analogy with the PQSCHA expression (\ref{e.xi}), but the
effective temperature $t_{\rm{eff}}^{\rm{H}}$ is defined through a
renormalization constant which is again a function of the
pure-quantum fluctuations only! Such remarkable and unexpected
feature of the cutoff-corrected field-theory prediction, suggested us
to substitute the perturbative expression $\xi_{\rm{3l}}^{\rm{cl}}$
with the exactly known classical $\xi_{\rm{cl}}$ thus getting the
results presented in Fig.~\ref{f.Fig23beard}.

It is thus made clear that the main features of the quantum
correlation length at intermediate temperatures are due to
essentially classical non-linear effects, which cannot be taken into
account by perturbative approaches. Moreover, the effective exchange
constant which defines the effective temperature
$t_{\rm{eff}}^{\rm{H}}$ is seen to depend on the same pure-quantum
renormalization coefficients defined by the PQSCHA, according to an
expression which is very similar (equal) to that found by the latter
approach in its standard (low-$T$) version: the behavior of $\xi$ in
the full temperature and spin-value ranges can thus be quantitatively
described by Eq. (\ref{e.xi}) without any adjustable fitting
parameter.

The results obtained by the PQSCHA about the correlation length and
other static quantities can also represent the needed information to
be inserted within other frameworks, like {\em mode-coupling} theory,
to interpret experiments probing dynamic quantities, like nuclear
magnetic resonance (NMR): an example of the successful combination of
PQSCHA and mode-coupling theory is given in
Ref.~\onlinecite{CCCMRTV2000prl}.

\section{Two-dimensional anisotropic Heisenberg model}
\label{s.anisotropic}

While the theoretical debate mentioned in the previous Section has
been mainly dedicated to the isotropic 2D QHAF, real compounds are
not actually well described by the isotropic model when the
temperature is low: indeed, the Mermin-Wagner
theorem~\cite{MerminW1966} states that a finite-temperature
transition cannot occur in the 2D isotropic QHAF, while the
experimental evidence of a transition suggests that 3D correlations
and anisotropy effects, as well as a combination of both, must be
considered. Easy-axis (EA) or easy-plane (EP) anisotropies turn out
to be fundamental in the analysis of the critical behavior.

\subsection{2D antiferromagnet with easy-axis anisotropy}

Several works (see Ref.~\onlinecite{Kanamori1963} for a review) have
shown that many additional interaction mechanisms may be taken into
account by inserting proper anisotropy terms in the magnetic
Hamiltonian; in particular, the transition observed in
K$_2$NiF$_4$~\cite{BirgeneauGS1970} ($S\,{=}\,1$),
Rb$_2$FeF$_4$~\cite{BirgeneauGS1970} ($S\,{=}\,2$),
K$_2$MnF$_4$~\cite{BirgeneauGS1973},
Rb$_2$MnF$_4$~\cite{BirgeneauGS1970} ($S\,{=}\,5/2$), and others, is
seen to be possibly due to an easy-axis anisotropy. Such anisotropy
has been often described in the literature through an external
staggered magnetic field in order to allow for a qualitative
description of the experimental data. However, this choice lacks the
fundamental property of describing a genuine phase transition, as the
field explicitly breaks the symmetry and makes the model ordered at
all temperatures. To preserve the symmetry under inversion along the
easy-axis, it is actually appropriate to insert an exchange
anisotropy term in the spin Hamiltonian. Spontaneous symmetry
breaking manifests then itself as a phase transition between ordered
and disordered states. The EA-QHAF Hamiltonian reads then
\begin{equation}
 {\cal H}=\frac J2\sum_{\bm{i,d}}\Big[\mu\big(
 {S}_{\bm{i}}^x{S}_{\bm{i+d}}^x+
 {S}_{\bm{i}}^y{S}_{\bm{i+d}}^y\big)+
 {S}_{\bm{i}}^z{S}_{\bm{i+d}}^z\Big]
\label{e.EAQHAF}
\end{equation}
where ${\bm{i}}\,{=}\,(i_1,i_2)$ runs over the sites of a square
lattice, ${\bm{d}}$ connects each site to its four nearest neighbors,
$J\,{>}\,0$ is the exchange integral and $\mu$ is the easy-axis
anisotropy parameter ($0\,{\le}\,\mu\,{<}\,1$). Again,
$J\widetilde{S}^2\,{\equiv}\,J(S{+}1/2)^2$ sets the overall energy
scale and $t\,{=}\,T/J\widetilde{S}^2$ is the reduced temperature.
When $\mu\,{=}\,1$ the model reduces to the isotropic QHAF. Note that
a canonical transformation reversing the $x$ and $y$ spin components
one one sublattice is equivalent to setting $\mu\,{\to}\,{-\mu}$, so
that the physical properties of the model are even functions of
$\mu$. The $\mu\,{=}\,0$ case is called {\em Ising limit}, not to be
confused with the genuine Ising model~\cite{Onsager1944}, reproduced
by Eq.~(\ref{e.EAQHAF}) with $\mu\,{=}\,0$ and $S\,{=}\,1/2$. Despite
being a very particular case of Eq.~(\ref{e.EAQHAF}), the 2D Ising
model on the square lattice is a fundamental point of reference for
the study of the thermodynamic properties of the EA-QHAF. A
renormalization-group analysis~\cite{BanderM1988} of the classical
model predicted the occurrence of an Ising-like transition at a
finite temperature $t_{\rm{c}}^{\rm{cl}}(\mu)$ of the order of unity
for any value of $\mu$, no matter how near to the isotropic value
$\mu\,{=}\,1$; this analysis received the support of several Monte
Carlo
simulations~\cite{PattersonJ1971,BinderL1976,SerenaGL1993,GouveaWLPKM1999}.

As for the quantum case, up to a few years ago no information was
available about the value of the critical temperature
$t_{\rm{c}}(\mu,S)$ as a function of anisotropy and spin, save the
fact that $t_{\rm{c}}(0,1/2)\,{=}\,0.567J$ (Onsager
solution~\cite{Onsager1944}) and $t_{\rm{c}}(1,S)\,{=}\,0$ (isotropic
limit). As a consequence it was also uncertain whether or not the
small anisotropy ($1-\mu\,{\simeq}\,{10^{-2}}$) observed in real
compounds could be responsible of transitions occurring at critical
temperatures of the order of $J$, also accounting for the fact that
quantum fluctuations are expected to lower the critical temperature
with respect to the classical case.

Over the last few years our group developed a quantitative analysis
of several thermodynamic properties of the model, by means of the
effective Hamiltonian method~\cite{CTVV1992pra,CGTVV1995jpcm} for
spin $S\ge{1}$ and by means of quantum Monte Carlo
simulations~\cite{CRTVV2003prb} in the case $S\,{=}\,1/2$.

\begin{figure}
\includegraphics[width=85mm,angle=0]{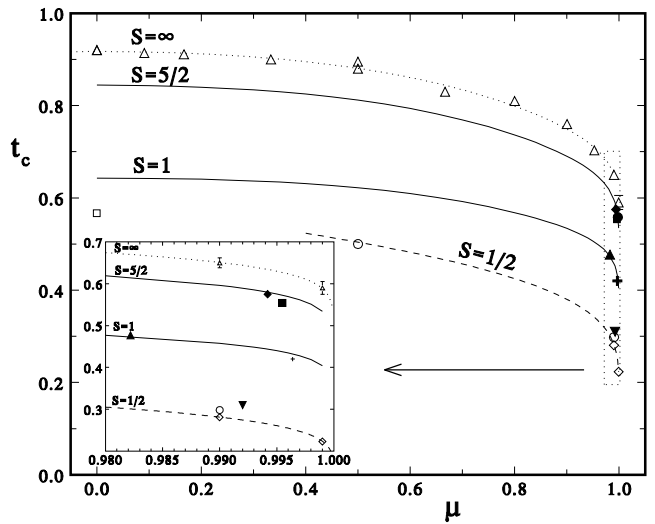}
\caption{
Critical temperature vs anisotropy $\mu$ for the 2D easy-axis
antiferromagnet. Dotted line: the fit of $t_{\rm{c}}^{\rm{cl}}(\mu)$
build up from classical MC
data~\cite{GouveaWLPKM1999,SerenaGL1993,CRTVV2001epjb} (open
triangles). Solid lines: PQSCHA result for $S\,{=}\,1$ and $5/2$.
Quantum MC data (open circles~\cite{Ding1990} and
diamonds~\cite{CRTVV2003prb}) for $S\,{=}\,1/2$, asymptotically
described by $t_{\rm{c}}(\mu)\,{=}\,2.49/\ln[70/(1{-}\mu)]$ (dashed
line), and exact result for the Ising model, $\mu\,{=}\,0$ (open
square). Experimental data for the compounds YBa$_2$Cu$_3$O$_6.1$
(down triangle~\cite{RossatMignodEtal1991}), K$_2$NiF$_4$
(cross~\cite{deWijnWW1973,SkalyoEtal1969}), Rb$_2$Ni F$_4$ (up
triangle~\cite{NagataT1974}), Rb$_2$MnCl$_4$
(circle~\cite{SchroederEtal1980}), Rb$_2$MnF$_4$
(diamond~\cite{deWijnWW1973,CowleyEtal1977}). In the inset the region
of weak anisotropy is enlarged. }
\label{f.ea.phd}
\end{figure}

The effective
Hamiltonian~\cite{CTRVV2000prb,CTVRV2001jmmm,CRTVV2001epjb} for the
EA-QHAF is expressed for classical spins as
\begin{equation}
 \frac{{\cal H}_{\rm{eff}}}{J\widetilde{S}^2}
 =\frac{j_{\rm{eff}}}2\sum_{\bm{i,d}}\big[
 \mu_{\rm{eff}}\big(s_{\bm{i}}^x s_{\bm{i+d}}^x+
 s_{\bm{i}}^y s_{\bm{i+d}}^y\big)+
 s_{\bm{i}}^z s_{\bm{i+d}}^z\big]~,
\label{e.EAQHAFeff}
\end{equation}
and shows a weaker renormalized exchange
$j_{\rm{eff}}(t,\mu)\,{<}\,1$ and easy-axis anisotropy
$\mu_{\rm{eff}}(t,\mu)\,{>}\,\mu$, besides an additional free-energy
term that is not reported. By means of ${\cal{H}}_{\rm{eff}}$ a
series of thermodynamic quantities were
studied~\cite{CTRVV2000prb,CRTVV2003jap1}: internal energy, specific
heat, staggered magnetization, staggered correlation function,
staggered correlation length, staggered susceptibility. This required
extensive classical Monte Carlo simulations, as varying the
temperature gives an effective system with different effective
anisotropy $\mu_{\rm{eff}}(t,\mu)$. The quantum phase diagram
reported in Fig.~\ref{f.ea.phd} could be built
up~\cite{CTVRV2000bjp,CRTVV2001epjb} in a simpler way starting from
the knowledge of the classical one and using a relation that follows
from the form of Eq.~(\ref{e.EAQHAFeff}),
\begin{equation}
 t_{\rm{c}}(\mu,S)=j_{\rm{eff}}(t,\mu,S)
 ~t_{\rm{c}}^{\rm{cl}}\big(\mu_{\rm{eff}}(t,\mu,S)\big) ~.
\end{equation}

In the region of very weak anisotropy, which is the most important in
view of the characterization of experimentally accessible materials,
we verified that the Ising-like transition temperature decreases very
slowly (logarithmically) towards its vanishing value in the isotropic
limit, so that $t_{\rm{c}}$ remains substantially of the order of
unity.

\begin{figure}
\includegraphics[width=85mm,angle=0]{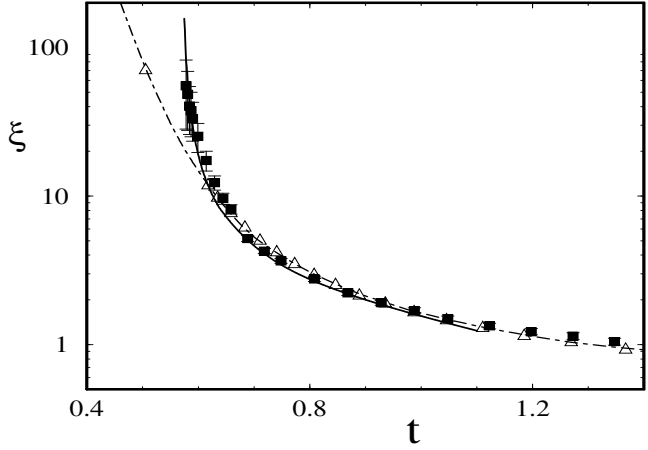}
\caption{
Correlation length vs $t$ for $S\,{=}\,5/2$, $\mu\,{=}\,0.9942$ (full
curve) and $\mu\,{=}\,1$ (isotropic, dash-dotted curve); the symbols
are neutron scattering data~\cite{LeeEtal1998} for Rb$_2$MnF$_4$. The
triangles are quantum Monte Carlo data~\cite{Beard2000} for the
isotropic model.}
\label{f.ea.Rb2MnF4}
\end{figure}

As a sample of the various results that were obtained, we report in
Fig.~\ref{f.ea.Rb2MnF4} the comparison~\cite{CTRVV2000prb} with the
experimental data~\cite{LeeEtal1998} for the correlation length of
the $S\,{=}\,5/2$ magnet Rb$_2$MnF$_4$, that results quite well
described by the anisotropic model with $J\,{=}\,7.42$~K and
$\mu\,{=}\,0.9942$. Rb$_2$MnF$_4$ is known to behave as a 2D magnet
both above and below the observed transition~\cite{BirgeneauGS1970},
so that the critical behavior is not contaminated by the onset of 3D
order and a clean characterization of the transition is possible. In
Ref.~\onlinecite{CTRVV2000prb} we have compared our theoretical
results also with the neutron scattering experimental data for the
staggered magnetization, staggered susceptibility and correlation
length of Rb$_2$MnF$_4$ and found an excellent agreement both for the
overall temperature behavior and for the value of the critical
temperature, that perfectly coincides with the one deriving from the
experimental analysis, $T_{\rm{c}}\,{=}\,38.4$~K (i.e.,
$t_{\rm{c}}\,{=}\,0.575$).

\begin{figure}
\includegraphics[width=85mm,angle=0]{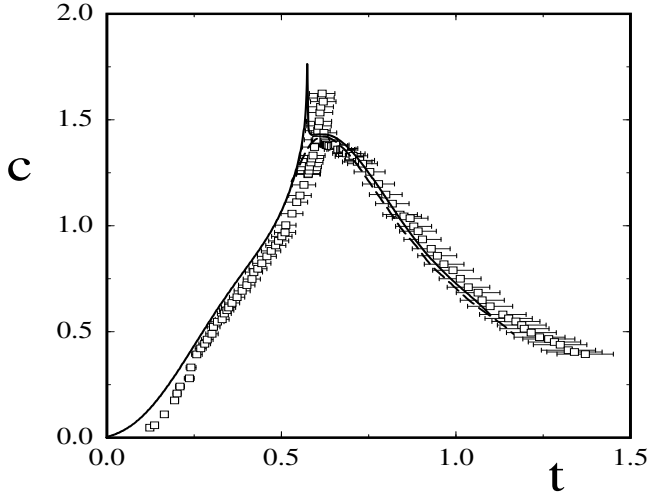}
\caption{
Specific heat {vs} $t\,{=}\,T/J\widetilde{S}^2$, for $S\,{=}\,5/2$.
Mn-f-2U experiments (squares)~\cite{TakedaEtal2001}, EA-QHAF with
$\mu\,{=}\,0.9942$ (solid line)~\cite{CRTVV2001epjb,CRTVV2003jap1},
isotropic QHAF (dashed line). Note that the correct anisotropy for
this compound is estimated to be $\mu\,{=}\,0.981$. Error bars are
due to the experimental uncertainty on $J$ for Mn-f-2U.}
\label{f.ea.MnF2U}
\end{figure}

Another quantity that shows a signature of the an\-isotropy and of
the Ising transition is the specific heat: in Fig.~\ref{f.ea.MnF2U} a
comparison with experimental data is shown in the case of the
$S\,{=}\,5/2$ compound Mn-formate di-Urea~\cite{CRTVV2003jap1}, whose
anisotropy can be estimated from the sole knowledge of the exchange
integral and of the measured transition temperature to be
$\mu\,{=}\,0.981$. The comparison reveals the existence of a
crossover from a high-temperature 2D-Heisenberg regime to a critical
2D-Ising regime that triggers the observed~\cite{TakedaEtal1989} 3D
phase transition at $T_{_{\rm{N}}}\,{=}\,3.77$~K.

\begin{figure}
\includegraphics[width=85mm,angle=0]{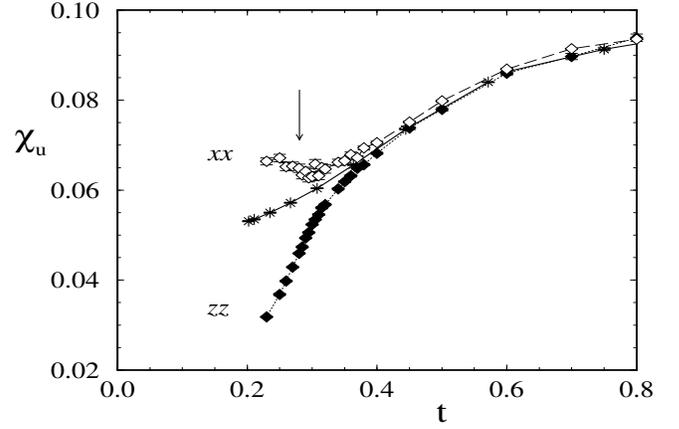}
\caption{
Uniform susceptibility of the EA model for $\mu\,{=}\,0.99$ from
quantum Monte Carlo simulations~\cite{CRTVV2003prb}. Full diamonds:
longitudinal branch; open diamonds: transverse branch; stars: data
for the isotropic model~\cite{KimT1998}. The lines are guides for the
eye. The arrow indicates the estimated critical temperature.}
\label{f.ea.chis}
\end{figure}

Finally, for the strongest quantum case, $S\,{=}\,1/2$, we have used
the continuous-time quantum Monte Carlo method based on the loop
algorithm~\cite{CRTVV2003prb}. The general outcome of the numerical
simulations is that the thermodynamics of 2D quantum antiferromagnets
is extremely sensitive to the presence of weak easy-axis anisotropies
of the order of those of real compounds. For instance, in
Fig.~\ref{f.ea.chis} it is shown that for $\mu\,{=}\,0.99$ the
uniform susceptibility, which is a noncritical quantity, undoubtedly
shows a characteristic anisotropic behavior with a different
temperature dependence of the transverse and longitudinal branches:
the former displays a minimum and the latter monotonically goes to
zero, as expected for an EA antiferromagnet. This behavior results
from the anisotropy-induced spin ordering, that makes the system more
sensitive to the application of a transverse magnetic field, rather
than of a longitudinal one. Both the minimum of the in-plane
component and the decrease of the longitudinal one are close to the
transition, a feature also peculiar to the Ising model. Results for
the critical temperature at $S\,{=}\,1/2$ are already included in
Fig.~\ref{f.ea.phd}.

\subsection{2D antiferromagnet with easy-plane anisotropy}

In the case of an easy-plane anisotropy the Mermin-Wagner theorem
still holds, so that no finite-temperature transition towards a phase
with a finite order parameter may occur. However, a
Berezinskii-Kosterlitz-Thouless (BKT)
transition~\cite{Berezinskii1970,KosterlitzT1973}, related with the
existence of vortex-like topological excitations, is known to
characterize the class of the easy-plane models and may occur at a
critical temperature $t_{\rm{c}}(\lambda,S)\,{>}\,0$. The reference
system for the easy-plane class is the {\em planar rotator model}, or
{\em $XY$ model}, defined in terms of two-component spins: above
$t_{\rm{c}}$ the system is disordered, with exponentially decaying
correlation functions; in the region $0\,{<}\,t\,{<}\,t_{\rm{c}}$ the
system is in a critical phase with vanishing magnetization and
power-law decaying correlators ({\em quasi}-long-range order); at
$t\,{=}\,0$ the magnetization gets a finite value and the system is
ordered.

The observation of clear signatures of BKT critical behavior in real
magnets is a controversial issue. However, it can explain the
properties of several layered
compounds~\cite{deJongh1990,Johnston1997,GrevenBEKMS1995,SuhBMCJT1995}
whose high-temperature phase can be described by a purely 2D
Heisenberg Hamiltonian, with an exchange interaction often displaying
weak easy-plane (EP) anisotropies, on the order of
$10^{-2}\div10^{-4}$ times the dominant isotropic
coupling~\cite{deJongh1990,Johnston1997}. Symmetry and universality
arguments suggest that the EP anisotropy drives the system towards a
BKT behavior at finite temperature, and the enhanced intraplane
correlations trigger the transition to the observed 3D ordered state.
As a consequence, 2D critical behavior in close proximity of the
would-be BKT transition is masked by these 3D effects.

The EP-QHAF Hamiltonian reads
\begin{equation}
 {\cal H}=\frac J2\sum_{\bm{i,d}}
 \Big[{S}_{\bm{i}}^x{S}_{\bm{i+d}}^x
 + {S}_{\bm{i}}^y{S}_{\bm{i+d}}^y
 +\lambda\,{S}_{\bm{i}}^z{S}_{\bm{i+d}}^z\Big]~,
\label{e.EPQHAF}
\end{equation}
where $\lambda$ is the easy-plane anisotropy parameter
($0\,{\le}\,\lambda\,{<}\,1$). Again,
$J\widetilde{S}^2\,{\equiv}\,J(S{+}1/2)^2$ sets the overall energy
scale and $t\,{=}\,T/J\widetilde{S}^2$ is the reduced temperature.
When $\lambda\,{=}\,1$ the model reduces to the isotropic QHAF. Note
that a canonical transformation reversing the $x$ and $y$ spin
components one one sublattice is equivalent to setting
$(J,\lambda)\,{\to}\,(-J,-\lambda)$, so that negative values of
$\lambda$ ($-1\,{<}\,\lambda\,{\le}\,0$) correspond to the easy-plane
ferromagnet. The $\lambda\,{=}\,0$ case is called {\em $XY$~model} or
XX0~{\em model}. However, at variance with the planar rotator model,
out-of-plane fluctuations are present both in the classical and in
the quantum EP models. Nevertheless, if $|\lambda|\,{<}\,1$, the
classical EP model still undergoes a BKT phase
transition~\cite{Khokhlachev1976}. MC simulations of the classical
systems~\cite{KawabataB1982,GerlingL1984,CTV1995prb,EvertzL1996}
confirm that the planar and the XXZ model share the same qualitative
behavior, but the value of the transition temperature of the planar
model~\cite{GuptaB1992} lowers by 22\,\% in the XX0 (i.e., with
$\lambda\,{=}\,0$) model~\cite{CTV1995prb,EvertzL1996}, as a
consequence of out-of-plane fluctuations. A renormalization group
calculation~\cite{Khokhlachev1976} predicts that the transition
temperature vanishes logarithmically as the isotropic limit
$|\lambda|\,{\to}\,{1}$ is approached, and this was also verified in
classical MC simulations~\cite{CTV1995prb}.
Experiments~\cite{GrevenBEKMS1995,NakajimaYHEGB1995} and quantum MC
simulations~\cite{DingM1990-92} indicated that the qualitative
behavior of the BKT transition is preserved in the quantum system,
with only quantitative modifications of the critical parameters due
to the quantum fluctuations.

We applied the effective Hamiltonian
formalism~\cite{CTVV1992pra,CGTVV1995jpcm} to the EP-QHAF, finding
that it was necessary~\cite{CCTVV1998phd,CRTVV2001epjb} to resort to
the Villain or to the Holstein-Primakoff transformation, depending on
the anisotropy being strong or weak, respectively. While the above
approach gives reliable results (with a smooth enough connection at
intermediate anisotropy) for spin $S\ge{1}$, we adopted quantum Monte
Carlo
simulations~\cite{CRTVV2003prb,CRVV2003prl,CRTVV2003jap2,CRTVV2003jap1,CRTVV2004jmmm}
in the case $S\,{=}\,1/2$. The effective
Hamiltonian~\cite{CTVV1995prb,CCTVV1998phd,CRTVV2001epjb} for the
EP-QHAF, in terms of classical spins, takes the form
\begin{equation}
 \frac{{\cal H}_{\rm{eff}}}{J\widetilde{S}^2}
 =\frac{j_{\rm{eff}}}2\sum_{\bm{i,d}}\big[
 s_{\bm{i}}^x s_{\bm{i+d}}^x+
 s_{\bm{i}}^y s_{\bm{i+d}}^y+
 \lambda_{\rm{eff}} s_{\bm{i}}^z s_{\bm{i+d}}^z\big]~,
\label{e.EPQHAFeff}
\end{equation}
and displays a weaker renormalized exchange
$j_{\rm{eff}}(t,\lambda)\,{<}\,J$ and easy-plane anisotropy
$\lambda_{\rm{eff}}(t,\lambda)\,{>}\,\lambda$ (an additional
free-energy term is not reported). In analogy to the EA case, the BKT
transition temperature can be obtained by renormalization of the
classical one using the self-consistent relation
\begin{equation}
 t_{\rm{c}}(\lambda,S)=j_{\rm{eff}}(t,\lambda,S)
 ~t_{\rm{c}}^{\rm{cl}}\big(\lambda_{\rm{eff}}(t,\lambda,S)\big) ~.
\end{equation}

\begin{figure}
\includegraphics[width=85mm,angle=0]{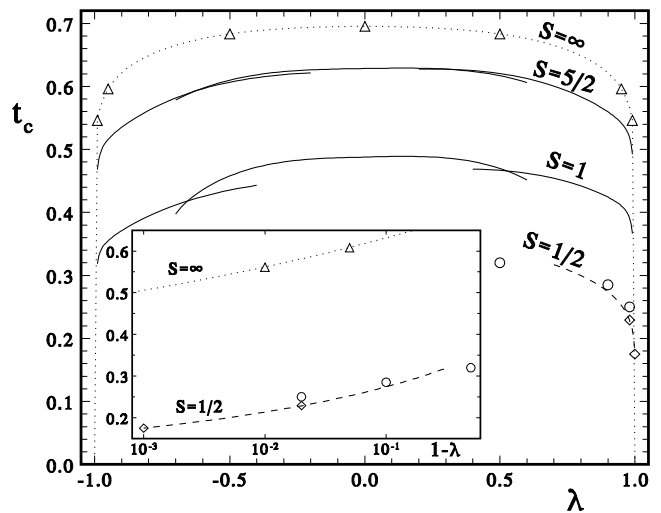}
\caption{
BKT critical temperature vs anisotropy $\lambda$ for the easy-plane
model with $S\,{=}\,1/2$, $1$, $5/2$, $\infty$. The triangles report
the classical ($S\,{=}\,\infty$) MC data~\cite{CTV1995prb} used to
construct the curve for $t_{\rm{c}}^{\rm{cl}}(\lambda)$ and hence the
renormalized curves for $S\,{=}\,1$ and $5/2$: the curves obtained
through the Holstein-Primakoff ($\lambda\,{\lesssim}\,0.5$) and the
Villain ($\lambda\gtrsim0.5$) spin-boson transformation are seen to
connect in a fairly smooth way. The diamonds are our QMC
data~\cite{CRTVV2003prb} for $S\,{=}\,1/2$, while the circles are
earlier QMC results~\cite{DingM1990-92}. The inset expands the nearly
isotropic region, in which the expected logarithmic behavior
$t_{\rm{c}}(\lambda)\,{\sim}\,(a-\ln|1{-}\lambda|)^{-1}$ is fitted by
the dashed curve.}
\label{f.ep.phd}
\end{figure}

In Fig.~\ref{f.ep.phd} the phase diagram of the EP-QHAF is reported,
including the QMC results for $S\,{=}\,1/2$. It is seen that the BKT
transition temperature stays large (i.e., comparable to the exchange
constant) also for very weak EP anisotropy.

\begin{figure}
\includegraphics[width=85mm,angle=0]{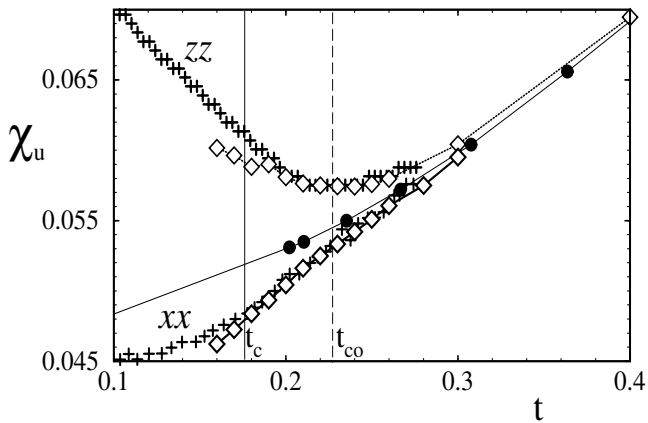}
 \caption{
Out-of-plane ($zz$) and in-plane ($xx$) uniform susceptibility
$\chi_{\rm{u}}$ for $\lambda\,{=}\,0.999$. Diamonds are our QMC
results, crosses are experimental
data~\cite{VakninEtal1990,VakninEtal1997} for Sr$_2$CuO$_2$Cl$_2$.
The circles report the result for the isotropic QHAF. The vertical
lines mark the 3D transition temperature $t_{_{\rm{N}}}\,{=}\,0.176$
and the crossover temperature $t_{\rm{co}}\,{\simeq}\,0.227$.}
\label{f.SCOC}
\end{figure}

However, as explained above, the problem of detecting the incipient
BKT transition requires to look for signatures of XY behavior in a
region above the transition. We have shown~\cite{CRVV2003prl} that a
suitable quantity is the uniform susceptibility
$\chi^{\alpha\alpha}_{\rm{u}}$, which has in-plane
($\alpha\,{=}\,x,y$) and out-of-plane ($\alpha\,{=}\,z$) components
and is noncritical, i.e., it does not show singularities at
$t_{\rm{c}}$. Fig.~\ref{f.SCOC} shows indeed that
$\chi^{zz}_{\rm{u}}$ deviates from the isotropic $\chi_{\rm{u}}$ and
displays a minimum. A similar feature is also present in other
quantities~\cite{CRTVV2003prb,CRVV2003prl} and occurs around the
temperature $t_{\rm{co}}(\lambda)$ that can be generally defined as
the minimum of $\chi^{zz}_{\rm{u}}(t,\lambda)$. The pronounced
deviation of $\chi^{zz}_{\rm{u}}$ from the isotropic behavior is due
to a simple statistical reason. The uniform susceptibility arises
from {\em spin canting}: two antiferromagnetically coupled spins in
an infinitesimal magnetic field $\bm{h}$ minimize their energy when
they lie almost orthogonal to $\bm{h}$ and slightly cant in its
direction, thus giving a linear response; if they are locally
parallel to $\bm{h}$ the response is instead negligible. When for
$t\,{\lesssim}\,t_{\rm{co}}$ the anisotropy becomes effective, the
fraction of spins aligned in the EP rapidly increases compared to
that of the isotropic case ($\sim{2/3}$) and the response to a field
along $z$ (i.e., $\chi^{zz}_{\rm{u}}$) is proportionally larger.

The layered cuprate Sr$_2$CuO$_2$Cl$_2$ is a good
realization~\cite{GrevenBEKMS1995} of the EP-QHAF model, with
$J\,{=}\,1450$\,K; considering the spin-wave gap renormalization, its
bare anisotropy is estimated to be $\Delta\,{\simeq}\,0.0014$.
Experimental data for the uniform susceptibility of this
compound~\cite{VakninEtal1990,VakninEtal1997}) are reported in
Fig.~\ref{f.SCOC}. They excellently compare with our results at
$\lambda\,{=}\,0.999$: the position of the minimum of
$\chi_{\rm{u}}^{zz}$ gives $t_{\rm{co}}\,{=}\,0.227(15)$. Close to
the critical region the experimental data are affected by the 3D
nature of the ordering of the real magnet: the N\'eel transition is
observed at $t_{_{\rm{N}}}\,{=}\,0.176(10)$ and well compares with
the theoretical estimate
$t_{\rm{c}}(\lambda{=}0.0014)\,{=}\,0.179(10)$, confirming that 3D
ordering is induced by the incipient intra-layer BKT transition.

It is worthwhile to mention that on the {\em triangular} lattice the
easy-plane antiferromagnet has very peculiar behavior, already at the
classical level, due to the frustration effect of accommodating three
antiferromagnetic spins on a plaquette. Indeed the minimum energy
corresponds to a configuration with the three sublattices aligned in
the easy plane at equal angles $2\pi/3$. As clockwise and
counterclockwise plaquette vorticities are possible, this
configuration is twofold degenerate and there is chiral symmetry,
which corresponds to an Ising-like order parameter. Therefore both a
BKT and an Ising transition coexist in the system. We have studied
the triangular antiferromagnet both in the
classical~\cite{CVCT1998prb} and in the quantum~\cite{CCTVV1999prb}
case, constructing the phase diagram for varying anisotropy and
showing that the transitions occur at slightly different
temperatures.

\subsection{2D antiferromagnet in an applied Zeeman field}

An interesting behavior is shown by the 2D Heisenberg antiferromagnet
when a magnetic field is applied. Indeed, a frustration phenomenon
occurs, as antiferromagnetism tends to antialign spins while the
field tends to align them with itself: in the classical
system~\cite{LandauB1981} it appears that the minimum-energy
configuration is the one with the spins almost orthogonal to the
field and canted in its direction. Therefore, provided the field is
not strong enough to overcome the exchange and to saturate the
magnetization, it acts as an effective easy-plane anisotropy and one
expects to observe BKT behavior. Remarkably, as this can be induced
in a real quasi-2D antiferromagnetic system by means of an applied
field, the strength of the effective anisotropy is in this case {\em
tunable}. Even though 2D criticality just acts as a trigger for 3D
ordering, by observing that the critical temperature follows the
predicted behavior upon the field, an experiment could realize an
objective observation of genuine 2D behavior.

The 2D QHAF in a uniform magnetic field is described by the
Hamiltonian
\begin{equation}
 {\cal H} = \frac{J}{2} \sum_{\bm{i},\bm{d}}
 {\bm{S}}_{\bm{i}}{\cdot}{\bm{S}}_{\bm{i}+\bm{d}}
 - g\mu_{_{\rm{B}}} H \sum_{\bm{i}}
 {S}^z_{\bm{i}}
\label{e.2DQHAF}
\end{equation}
where $H$ is the applied Zeeman field, $\mu_{_{\rm{B}}}$ the Bohr
magneton, and $g$ the gyromagnetic ratio.

\begin{figure}
\includegraphics[width=85mm,angle=0]{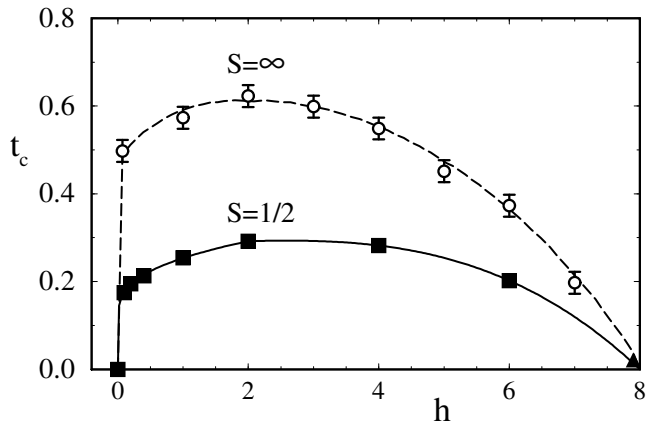}
\caption{Phase diagram of the $S\,{=}\,\frac12$
2D\,QHAF in a magnetic field. Open symbols refer to the classical
limit of the model~\cite{LandauB1981}; the
triangle~\cite{Syljuasen2000} and the squares~\cite{CRVV2003prb} are
QMC results.}
\label{f.h_tc}
\end{figure}

We have studied~\cite{CRVV2003prb,CRVV2004jmmm} the $S\,{=}\,\frac12$
2D QHAF in an uniform magnetic field by means of the QMC method based
on the worm algorithm. Our results confirmed that an arbitrarily
small field is able to induce a BKT transition and an extended $XY$
phase above it, as in the case of an easy-plane exchange anisotropy.
The field-induced $XY$ behavior becomes more and more marked for
increasing fields, while for strong fields the antiferromagnetic
behavior along the field axis is nearly washed out, so that the
system behaves as a planar rotator model with antiferromagnetism
surviving in the orthogonal plane only; the BKT critical temperature,
as reported in Fig~\ref{f.h_tc} (where $t\,{\equiv}\,{T/J}$ and
$h\,{\equiv}\,2{g}\mu_{_{\rm{B}}}H/J$), vanishes as the field reaches
the saturation value $h_{\rm{c}}$ and the effective rotator length
goes to zero. We have therefore shown that the model in a moderately
strong field represents an ideal realization of the $XY$ model and
that $XY$ behavior can be detected by measuring standard non-critical
quantities, as the specific heat or the induced magnetization; an
experimental realization of the $XY$ model in purely magnetic systems
and a systematic investigation of the dynamics of vortex/antivortex
excitations is therefore possible.

\section{Frustration in the 2D quantum $J_1$-$J_2$ Heisenberg model}
\label{s.J1J2}

The study of frustrated quantum spin systems is one of the most
challenging and exciting topics in theoretical magnetism because of
the possible existence of a non magnetic zero temperature phase. A
very extensively investigated, yet largely debated model is the
so-called $J_1$-$J_2$ Heisenberg model with competing
antiferromagnetic couplings ($J_1,J_2\,{>}\,0$) between
nearest-neighbors (nn) and next-nearest-neighbors (nnn)
\begin{equation}
{\cal H} = J_1 \sum_{\rm{nn}}{\bm{S}}_{\bm{i}}{\cdot}{\bm{S}}_{\bm
j}+ J_2
\sum_{\rm{nnn}}{\bm{S}}_{\bm{i}}{\cdot}{\bm{S}}_{\bm j}~, \label{e.j1j2}
\end{equation}
where the spin operators are defined on a periodic lattice with
$N\,{=}\,L\times{L}$ sites; hereafter $\alpha\,{=}\,J_2/J_1$ defines
the frustration ratio.

In the classical limit ($S\,{\to}\,\infty$), the minimum energy
configuration has conventional N\'eel order with magnetic wave vector
${\bm{Q}}\,{=}\,(\pi,\pi)$ for $\alpha\,{<}\,0.5$. Instead, for
$\alpha\,{>}\,0.5$, the antiferromagnetic order is established
independently on the two sublattices, with the two staggered
magnetizations free to rotate with respect to each other. One of the
two families of {\em collinear} states, with pitch vectors
${\bm{Q}}\,{=}\,(\pi,0)$ or $(0,\pi)$, are selected by an
order-by-disorder mechanism as soon as thermal or quantum
fluctuations are taken into account. As a result, for
$\alpha\,{>}\,0.5$ the classical ground state breaks not only the
spin rotational and translational invariance of the Hamiltonian -- as
the conventional N\'eel phase -- but also its invariance under
$\pi/2$ lattice rotations, the resulting degeneracy corresponding to
the group $O(3)\times Z_2$. Remarkably, the additional {\em discrete}
$Z_2$ symmetry can in principle be broken at finite temperatures
without violating the Mermin-Wagner theorem. On this basis, in a
seminal paper~\cite{ChandraCL1999}, Chandra, Coleman, and Larkin
(CCL) proposed that the 2D $J_1$-$J_2$ model could sustain an Ising
phase transition at finite-temperature, with an order parameter
directly related to the $Z_2$ degree of freedom induced by
frustration. They also provided quantitative estimates of the
critical temperatures in the large-$\alpha$ limit for both the
classical and the quantum cases.

This transition in the classical model has been established by an
extensive Monte Carlo~\cite{WeberCMBEM2003}. In the quantum case, the
occurrence of a low-temperature phase with a discrete broken symmetry
has been subject of debates in connection with the discovery of three
vanadate compounds (Li$_2$VOSiO$_4$, Li$_2$VOGeO$_4$, and VOMoO$_4$)
whose relevant magnetic interactions involve nearest and next-nearest
spin-$1/2$ $V^{4+}$ ions on weakly coupled stacked planes. In
particular, NMR and $\mu$SR measurements on
Li$_2$VOSiO$_4$~\cite{MelziEtal2001-02} indicate the occurrence of a
transition to a low-temperature phase with collinear order at
$T_{_{\rm{N}}}\,{\simeq}\,2.8$ K. However in the experiments with
vanadate compounds, structural distortions, interlayer and anisotropy
effects are likely to come into play~\cite{CarrettaEtal2002}, and on
the other hand the theoretical investigation cannot rely on the
insight provided by quantum Monte Carlo methods as their reliability
in presence of frustration is strongly limited (see the review
articles in Ref.~\onlinecite{KotovOSZ2000,Capriotti2001}).

A complete study of the thermodynamic properties of the the quantum
$J_1$-$J_2$ model in its collinear phase has been pursued within the
PQSCHA scheme~\cite{CGTVV1995jpcm} described in
Section~\ref{s.isotropic}, by which the thermodynamics is rephrased
in terms of a classical effective Hamiltonian with renormalized
parameters depending on the spin value, temperature, and frustration.
It is possible to show that, to $O(1/S)$, the effective Hamiltonian
can be recast in a form preserving all the symmetries of the original
model, and that reads (except for uniform terms):
\begin{equation}
{\cal H}^{\rm{eff}}=J_1^{\rm{eff}}\widetilde{S}^2 \sum_{\rm{nn}}
{\bm{s}}_{\bm{i}}{\cdot}{\bm{s}}_{\bm j}~+~ J_2^{\rm{eff}}
\widetilde{S}^2
\sum_{\rm{nnn}} {\bm{s}}_{\bm{i}}{\cdot} {\bm{s}}_{\bm j} ~,
\label{e.j1j2eff}
\end{equation}
where ${\bm{s}}_{\bm{i}}$ are classical vectors of length 1,
$\widetilde{S}\,{=}\,S\,{+}\,\frac12$ is the effective spin length,
and
$J^{\rm{eff}}_1\,{=}\,(\theta_x^2{+}\,\theta_y^2)\,\theta_2^2J_1/2$,
$J^{\rm{eff}}_2\,{=}\,\theta_2^4J_2$, are the quantum-renormalized
exchange integrals, with spin-, temperature- and
frustration-dependent renormalization parameters $\theta_x$,
$\theta_y$ and $\theta_2$.
\begin{figure}
\includegraphics[width=85mm,angle=0]{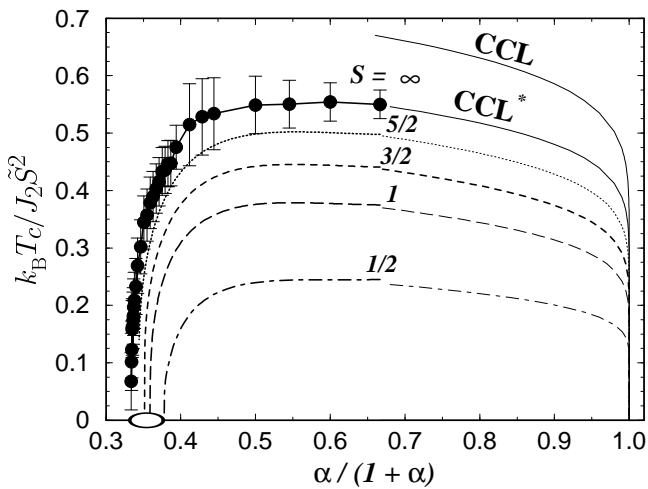}
 \caption{\label{f.tc}
Renormalized critical temperature of the CCL transition for various
values of the spin. Classical data ($\bullet$) are taken from
Ref.~\onlinecite{WeberCMBEM2003}. The solid lines on the right are
the CCL and the CCL$^*$ prediction for the classical and
$S\,{=}\,1/2$ case (see text). The ellipse on the horizontal axis
mark the non-magnetic (spin-liquid) phase between
$\alpha_{\rm{c}}(S\,{=}\,\infty)\,{=}\,0.5$ and
$\alpha_{\rm{c}}(S\,{=}\,1/2)\,{\simeq}\,{0.6}$. }
\end{figure}

The occurrence of the transition~\cite{ChandraCL1999} in the quantum
case can be directly addressed within our approach by calculating the
critical temperatures as functions of the spin and of the frustration
ratio~\cite{CFRT2003prl}. Using a simple scaling argument the
critical temperatures in the quantum case $T_{\rm{c}}(S,\alpha)$ can
be related to those of the classical model
$T^{(cl)}_{\rm{c}}(\alpha)$ through the following self-consistent
relation~\cite{CRTVV2001epjb,CCTVV1999prb}
\begin{equation}
 T_{\rm{c}}(S,\alpha) = j_1^{\rm{eff}}(T_{\rm{c}},S,\alpha)~
 T^{(cl)}_{\rm{c}}(\alpha^{\rm{eff}}(T_{\rm{c}},S,\alpha))~,
\label{e.tccl}
\end{equation}
where $j_1^{\rm{eff}}\,{=}\,J_1^{\rm{eff}}\widetilde S^2/J_1$ and
$\alpha^{\rm{eff}}\,{=}\,J_2^{\rm{eff}}/J_1^{\rm{eff}}$. The
classical transition temperature, $T^{(cl)}_{\rm{c}}(\alpha)$ is
accurately known through extensive MC simulations for
$\alpha\,{\le}\,2$; it vanishes for $\alpha\,{\to}\,1/2$ and grows
more or less linearly for $\alpha\,{>}\,1$.

The behavior of the transition temperature versus the frustration
ration is plotted in Fig.~\ref{f.tc} for different values of the spin
length. In order to represent the whole interval of
$\alpha\,{\in}\,[1/2,\infty)$ in Fig.~\ref{f.tc} we have plotted both
the MC and the CCL estimates of the classical critical temperatures
as a function of $\alpha/(1+\alpha)$. The mismatch between the MC and
CCL predictions is a minor flaw that can be corrected by slightly
modifying CCL's criterion for the determination of the transition
temperature as explained in Ref.~\onlinecite{WeberCMBEM2003}.
Remarkably, while for large $\alpha$ the transition temperature
vanishes for $\alpha\,{\to}\,\infty$ for any value of the spin, in
the opposite limit the critical temperatures vanish approaching a
critical value $\alpha_{\rm{c}}\,{>}\,0.5$ that increases as $S$
decreases, thus confirming the existence of a non-magnetic phase in
the regime of high frustration. In particular for $S\,{=}\,1/2$,
$\alpha_{\rm{c}}\,{\simeq}\,0.6$ in agreement with the previous
estimates of the zero-temperature quantum critical
point~\cite{Capriotti2001}.

\section{Conclusions}

The activity in magnetism of the Condensed Matter Theory group in
Firenze~\cite{CMTG} stems from the early work on two-magnon Raman
scattering in the seventies, and has grown up in the years with the
collaboration of several scientists. In this paper we summarize the
relevant theoretical work that concerns antiferromagnetic models.
This activity has been mainly concentrated on low-dimensional systems
and has found one of its main motivations in the intent of
interpreting the data collected in experiments on real materials.
Among the prominent subjects, we reported about soliton-excitation
effects in one-dimensional systems, critical and near-to-critical
behaviors and phase transitions in two-dimensional ones. Besides
that, we also faced some intriguing theoretical problems where
fundamental aspects of Quantum Mechanics come into play, as, e.g.,
the ground state of antiferromagnetic chains with integer spin or the
possible quantum critical regime predicted from the field-theory
treatment of the two-dimensional antiferromagnetic Heisenberg model.

\end{document}